\documentclass[%
 reprint,
superscriptaddress,
 amsmath,amssymb,
 aps,
 prl,
 longbibliography,
]{revtex4-2}

\usepackage{graphicx}% Include figure files
\usepackage{dcolumn}% Align table columns on decimal point
\usepackage{bm}% bold math
\usepackage[colorlinks=true,allcolors=blue]{hyperref}%
\usepackage[flushleft]{threeparttable}
\usepackage{booktabs}
\usepackage{relsize}
\usepackage{csquotes}

\begin{document}

\title{Prediction and Observation of Intermodulation Sidebands from Anharmonic Phonons in NaBr}
%\title{Ghost Phonons in an Anharmonic Crystal of NaBr}
%\title{Ghost Phonons in NaBr from Asymmetric Phonon Intermodulation}

\author{Y. Shen}
\affiliation{Department of Applied Physics and Materials Science, California Institute of Technology, Pasadena, California 91125, USA}
\author{C. N. Saunders}
\affiliation{Department of Applied Physics and Materials Science, California Institute of Technology, Pasadena, California 91125, USA}
\author{C. M. Bernal}
\affiliation{Department of Applied Physics and Materials Science, California Institute of Technology, Pasadena, California 91125, USA}
\author{D. L. Abernathy}
\affiliation{Neutron Scattering Division, Oak Ridge National Laboratory, Oak Ridge, Tennessee 37831, USA}
\author{T. J. Williams}
\affiliation{Neutron Scattering Division, Oak Ridge National Laboratory, Oak Ridge, Tennessee 37831, USA}
\author{M. E. Manley}
\affiliation{Materials Science and Technology Division, Oak Ridge National Laboratory, Oak Ridge, Tennessee 37831, USA}
\author{B. Fultz}
\email{btf@caltech.edu}
\affiliation{Department of Applied Physics and Materials Science, California Institute of Technology, Pasadena, California 91125, USA}

\date{\today}% It is always \today, today,


\begin{abstract}

A quantum Langevin model, similar to models from optomechanics,
was developed for phonons. It predicts intermodulation phonon sidebands (IPS) in anharmonic crystals.
\emph{Ab initio} calculations of anharmonic phonons in rocksalt NaBr
%based on anharmonic perturbation theory also 
showed  these  spectral features as ``many-body effects.''
Modern inelastic neutron scattering measurements on a 
crystal of NaBr at 300\,K
revealed diffuse intensity at high phonon energy from a 
predicted upper IPS. 
The transverse optical (TO) part of 
the new features originates from phonon intermodulation between the transverse acoustic (TA) and TO phonons. The 
longitudinal optical (LO) spectral features originate from 
three-phonon
coupling between the TA modes and the TO lattice modes. 
The partner lower IPS proves to be an ``intrinsic
localized mode.''
Interactions with the thermal bath broaden and redistribute the spectral weight of the IPS pair.
These sidebands
%should participate in  light-matter interactions, and 
are a probe of the anharmonicity and quantum noise of phonons in NaBr, and suggest novel interactions between photons and phonons. 
\end{abstract}

\maketitle

\section{Introduction \label{Introduction}}

Phonons, quantized excitations of vibrational modes in crystals, bear  resemblance to photons, quantized excitations of electromagnetic fields.
Both obey the Planck statistics of bosons, and their quintessential models have  similar  Hamiltonians,
$\mathcal{H}_0 = \hbar \omega ( a^{\dagger}a + \frac{1}{2})$,
where $\hbar \omega$ is the energy of an individual phonon or photon, and $a^{\dagger}a$ gives the number of them. 
Phonons and photons exist in different media, so their properties are explained differently. 
For phonons, harmonic equations of motion are formulated as eigenvalue problems 
that give   
dispersions of frequency versus wavevector, $\omega(\vec{k})$ \cite{Chaikin1995, Ziman2001}. 
The degrees of freedom in three dimensions  allow 
$3 \mathcal{R}$ dispersions, where $\mathcal{R}$ is the number of atoms in the translationally-periodic
unit cell.
The harmonic model is readily extended to a ``quasiharmonic model''
to account for how frequencies
shift with volume. 
Anharmonic models based on 
many-body perturbation theory \cite{Maradudin1962, Wallace1972, Grimvall1986} 
can account for
how phonon frequencies shift with temperature alone, and how finite phonon lifetimes originate
with interactions between phonons. 
Perturbation theory couples the phonon modes, but the $3 \mathcal{R}$ dispersions
are retained. 
To date, these $3 \mathcal{R}$ dispersions
have been consistent with experimental observations.
Exceptions are predictions \cite{Sievers1988, Sievers2013}
%NaI definitely cite this:
%Physical Review B 88, 104305 (2013)
%Thermally populated intrinsic localized modes in pure alkali halide crystals
%A.J. Sievers, M. Sato, J.B. Page, and T. R\"{o}ssler
and  experimental reports of 
intrinsic localized modes (ILM) \cite{Sato2004, Manley2009, Manley2011, Manley2014, Manley2019}.
There are different viewpoints about the experimental 
evidence for ILMs, however \cite{Kempa2014, Pailhes2014, Riviere2019}.

%Phonons, energy quanta in vibrational modes, are essential to condensed matter physics \cite{Chaikin1995, Ziman2001}.
%Harmonic models give the essence of 
%phonon dispersions of frequency versus wavevector, $\omega(\vec{k})$, 
%and the harmonic equations of motion are formulated as eigenvalue problems. 
%In three dimensions these give
%$3 \mathcal{R}$ dispersions, where $\mathcal{R}$ is the number of atoms in the
%unit cell.
%The harmonic model is readily extended to a ``quasiharmonic model''
%to account for how the dispersions
%shift with volume. 
%Anharmonic models based on 
%many-body perturbation theory \cite{Maradudin1962, Wallace1972, Grimvall1986} 
%are needed to account for
%how the dispersions shift with temperature alone, and how finite phonon lifetimes originate
%with phonon-phonon interactions. 
%Perturbation theory couples the phonon modes, but the $3 \mathcal{R}$ dispersions
%are retained. 

In the newer field of laser-cavity physics, a quantized mechanical motion is coupled to photons in a cavity. 
%A laser can be used to cool or heat the mechanical motion, which is described as a quantum harmonic oscillator \cite{Clerk2010, Kippenberg2008, Vahala2009, Chan2011, Benz2016, Riedinger2016, Renninger2018, Tavernarakis2018}.
When tuning the laser across the resonant frequency of the cavity, 
%can be used to realize the regenerative oscillation
cooling or heating of the mechanical system 
generates sidebands about the main resonance \cite{Clerk2010, Kippenberg2008, Vahala2009, Chan2011, Benz2016, Riedinger2016, Renninger2018, Tavernarakis2018}.
%Recent works on two dimensional system show the connections between photons and localized phonons \cite{Sullivan2017}.
Photon-phonon couplings in laser-cavity experiments
have similarities to 
anharmonic phonon-phonon couplings in crystals. 
The formal similarities motivate the question, 
``Do thermally-driven asymmetric sidebands exist in the phonon spectra of anharmonic crystals?''
%Advances in neutron scattering methods have allowed 
%us to observe and report a direct observation of a ``ghost'' phonon branch in NaBr, 
%which originates from anharmonic phonon intermodulation.
To date, there has been no experimental evidence for this.

Advances in the sensitivity of methods for
inelastic neutron scattering (INS) on single-crystals
motivated an examination of this question. 
After these experimental methods are described, this paper presents
a quantum Langevin model
for equilibrium phonon populations. 
INS data are presented on an anharmonic material, 
rocksalt NaBr,
revealing a new diffuse spectral band at 300\,K. 
The diffuse band  is predicted qualitatively by
ab initio calculations and 
perturbation theory with cubic perturbations to second order.
This ab initio method with perturbation theory is used for identifying the 
specific phonon energies and branches involved in
creating the new diffuse band. 
The quantum Langevin model is not limited to
small anharmonicity, however.
It successfully
explains the intensity and asymmetry of 
the intermodulation phonon 
sidebands (IPS) through the anharmonic
coupling of two phonons and their
interactions with a thermal  bath of other phonons
in the crystal.

%Here we reveal the possible phonon intermodulation mechanism %from the anharmonic phonon-phonon interactions, and report an evidence of new spectral features in NaBr at moderate temperatures, where spectral shapes modified by the quantum back action from a thermal phonon bath.
%Here we report new spectral features in NaBr at moderate temperatures, and show how they are intermodulation phonon sidebands, with spectral shapes modified by the quantum back action from a thermal phonon bath. 

\section{Experimental Measurements and {\it Ab Initio} Calculations}

\subsection{INS experiments}

The  measurements used a high-purity single crystal of NaBr. 
Crystal quality  was checked by X-ray and neutron diffraction. 
The single crystal of [001] orientation  was suspended in an aluminum holder, which was  mounted in a closed-cycle helium refrigerator for the 10\,K measurement, and in a low-background electrical resistance vacuum furnace for measurements at 300\,K. 

%
%Phonons in a single crystal of NaBr were measured by %inelastic neutron scattering (INS) experiments with the %time-of-flight spectrometer, ARCS \cite{Abernathy2012}, %at the Spallation Neutron Source at Oak Ridge National %Laboratory. 
%
The inelastic neutron scattering (INS) data  were acquired with the time-of-flight Wide Angular-Range Chopper Spectrometer, ARCS \cite{Abernathy2012}, at the Spallation Neutron Source at Oak Ridge National Laboratory, using neutrons with an incident energy,  $E_{\rm i}$, of 50\,meV. 
%(The Supplemental Information describes some
%additional measurements with a triple-axis spectrometer
%at a specific point in $Q$-space.)
The techniques and material were similar to those reported previously \cite{Shen2020}, but the previous
study had a problematic 
background at the precise energy transfers of interest here \footnote{The internal structure
of the ARCS spectrometer has radial, plate-like baffles of neutron absorbing material
that block stray scattering off  detector tubes from reaching other detectors. This works
well, except for detectors located approximately 180$^{\circ}$ across the detector array. 
Those neutrons, elastically 
scattering across the diameter of the instrument, arrive at a later time than the main elastic peak, appearing to be at an inelastic energy transfer of 80\% of  $E_{\rm i}$. 
In the prior dataset of $E_{\rm i} = 30$\,meV, this artifact appears at 24\,meV, 
overlapping the
ghost modes. For the present dataset with $E_{\rm i} = 50$\,meV, its 40\,meV artifact is 
safely out of range, although the energy resolution of the instrument is
approximately $5/3$ times larger than the prior dataset.}.
We therefore acquired an entirely new dataset with $E_{\rm i} = 50$\,meV. 

For each measurement, time-of-flight neutron data were collected from 201 rotations of the crystal in increments of 0.5$^{\circ}$ about the vertical axis. Data reduction gave the 4D scattering function $S(\mathbf{Q}, \varepsilon)$, where $\mathbf{Q}$ is the 3D wave-vector of momentum transfer, and $\varepsilon$ is the phonon energy (from the neutron energy loss). Measurements were  performed to evaluate the background  from an empty can. To correct for nonlinearities of the ARCS instrument, offsets of the $q$-grid were corrected to first order by fitting a set of 76 \emph{in situ} Bragg diffractions, which were  transformed to their theoretical positions in the reciprocal space of the NaBr structure. The linear  transformation matrix had only a small deviation (less than 0.02) from the identity matrix, showing that the original data had  good quality and that linear corrections for $q$-offsets were adequate. 
After subtracting the empty-can background and removing multiphonon scattering with the incoherent approximation, the higher Brillouin zones  were folded back \cite{Kim2018, Shen2020} into an irreducible wedge in the first Brillouin zone to obtain the spectral intensities.
Further information about the ARCS background is given in the Supplemental Information.
%shown in Fig. \ref{fig: exp_vs_cal}a-b. 

%\subection{HB3 Triple-Axis Spectrometer}

The temperature dependence of the low-energy dynamics of NaBr was measured with higher resolution 
with the HB3 triple axis spectrometer at the High Flux Isotope Reactor (HFIR) of Oak Ridge National Laboratory.
Pyrolytic graphite PG(002) was used for both the monochromator and the analyzer. 
The spectrometer was operated with a filtered, fixed final neutron energy of 14.7\,meV with horizontal collimation 48’:40’:40’:120’. The NaBr crystal was mounted in a vacuum furnace with the (HHL) reflections in the scattering plane. Measurements were made in transverse geometry near (113) along $\vec{Q}$=$[H,H,3]$ at
temperatures from 300 to 723\,K.

\subsection{\emph{Ab initio} calculations}

All DFT calculations were performed with the VASP package using a plane-wave basis set \cite{Kresse1993, Kresse1994, Kresse1996, Kresse1996_2} with  projector augmented wave (PAW) pseudopotentials \cite{Kresse1999} and the Perdew-Burke-Ernzerhof (PBE) exchange correlation functional \cite{Perdew1996}.  
The Born effective charges and dielectric constants were obtained by DFT calculations in VASP \cite{Gajdos2006}. A correction for the non-analytical  term  of  the  long-ranged  electrostatics was performed in both quasiharmonic and anharmonic calculations \cite{Gonze1997}. All calculations used a kinetic-energy cutoff of 550\,eV, a 5$\times$5$\times$5 supercell of 250 atoms, and a 3$\times$3$\times$3 $k$-point grid.
The phonon self-energy was calculated with a 35$\times$35$\times$35 $q$-grid.
Calculations of phonons in the quasiharmonic approximation (QHA) used PHONOPY \cite{Togo2015}. 
The QHA method allows the frequencies and entropy of phonons to
vary with volume and Planck occupancy. 
It does not include thermal displacements of individual atoms off periodic sites,
as arise in molecular dyanmics, for example, and the QHA
was not accurate for predicting the thermal expansion of NaBr \cite{Shen2020}.

The stochastically-initialized temperature dependent effective potential method (sTDEP) \cite{Hellman2011, Hellman2013, Hellman2013_2} method was used to accelerate the traditional \emph{ab initio} molecular dynamics (AIMD) and calculate anharmonic phonon dispersions at finite temperatures. 
The method for NaBr was described previously  \cite{Shen2020}.
In short, the phonon frequencies were obtained from the dynamical matrix for the quadratic force constants, and then corrected by the real ($\Delta$) and imaginary ($\rm{i} \Gamma$) parts of the phonon self-energy from many-body theory \cite{Wallace1972, Grimvall1986}. 
The imaginary part, which gives phonon lifetime broadening,
was calculated with the third-order force constants,
\begin{widetext}
\begin{align}
    \label{eq: CubicSecondOrder}
    \Gamma_{\lambda}(\Omega)  =  \frac{\hbar\pi}{16}\sum_{\lambda'\lambda''}\left|\Phi_{\lambda\lambda'\lambda''}\right|^2\big{\{}&(n_{\lambda'}+n_{\lambda''}+1)\times \delta(\Omega-\omega_{\lambda'}-\omega_{\lambda''}) \nonumber \\
    +&(n_{\lambda'}-n_{\lambda''}) \times \left[\delta(\Omega-\omega_{\lambda'}+\omega_{\lambda''}) -\delta(\Omega+\omega_{\lambda'}-\omega_{\lambda''})\right]\big{\}} \; , 
\end{align}
\end{widetext}
where $\Omega =E/\hbar$) is the probing energy. The real part was obtained by a Kramers-Kronig transformation
\begin{align}
    \Delta(\Omega)=\mathcal{P}\int\frac{1}{\pi}\frac{\Gamma(\omega)}{\omega-\Omega}\mathrm{d}\omega \; .
\label{eq: Kramers-Kronig}
\end{align}
Equation \ref{eq: CubicSecondOrder} is a sum over all possible three-phonon interactions, where $\Phi_{\lambda\lambda'\lambda''}$ is the three-phonon matrix element obtained from the cubic force constants by  Fourier transformation, $n$ is the Bose-Einstein thermal occupation factor giving the number of phonons in each mode, and the delta functions conserve energy and momentum. Details were given in the supplemental materials in our previous work \cite{Shen2020}.

\section{Quantum Langevin Model \label{QuantumLangevin}}

We start with the Hamiltonian of three coupled phonons denoted as $j$, $j^\prime$, and $j^{\prime \prime}$,
\begin{align}
    \mathcal{H_{\rm sys}}= \ \mathcal{H}_0 +\hbar\eta \left(\hat{a}^\dagger_{j}+\hat{a}_{j} \right) \left(\hat{a}^\dagger_{j^\prime}+\hat{a}_{j^\prime}\right)\left(\hat{a}^\dagger_{j^{\prime\prime}}+ \hat{a}_{j^{\prime\prime}}\right) \; ,
    \label{eq: HamiltonianForLOGhostPhononsSimplified}
\end{align}
where $\mathcal{H}_0=\sum_{k=j,j^\prime,j^{\prime\prime}}\hbar \omega_k (\hat{a}^\dagger_k \hat{a}_k+\frac{1}{2}) $ is the Hamiltonian for  three uncoupled, independent oscillators, and $\eta$ parameterizes the coupling strength.
However, there is also a special case where only two types of phonons are involved in this interaction process. Taking $j^\prime = j^{\prime \prime}$ as an example, the system Hamiltonian is then
\begin{align}
    \mathcal{H_{\rm sys}}\,=\ \mathcal{H}_0+ \hbar\frac{\eta}{2}\left(\hat{a}^{\dagger} +\hat{a}\right)^2\left(\hat{b}^\dagger+\hat{b}\right) \;,
\end{align}
where now $\hat{a}$ denotes the composite phonon mode
with $j^{\prime}=j^{\prime\prime}$, 
$\hat{b}$ denotes the $j$ mode, and the $1/2$ is added for later convenience. Confining our interest to terms under the rotating wave approximation (RWA) in quantum optics, we eliminate the terms $aab^{\dagger}$ and $a^{\dagger}a^{\dagger}b$ (and $aab$ and $a^{\dagger}a^{\dagger}b^{\dagger}$ that do not conserve energy)
\begin{align}
    \mathcal{H_{\rm sys}}\,=\ \mathcal{H}_0+
    \hbar\frac{\eta}{2}\left(\hat{a}^{\dagger}\hat{a}+\hat{a}\hat{a}^{\dagger}\right)\left(\hat{b}^\dagger+\hat{b}\right) \;.
\end{align}

\begin{figure}
    \centering
    \includegraphics[width=0.95\linewidth]{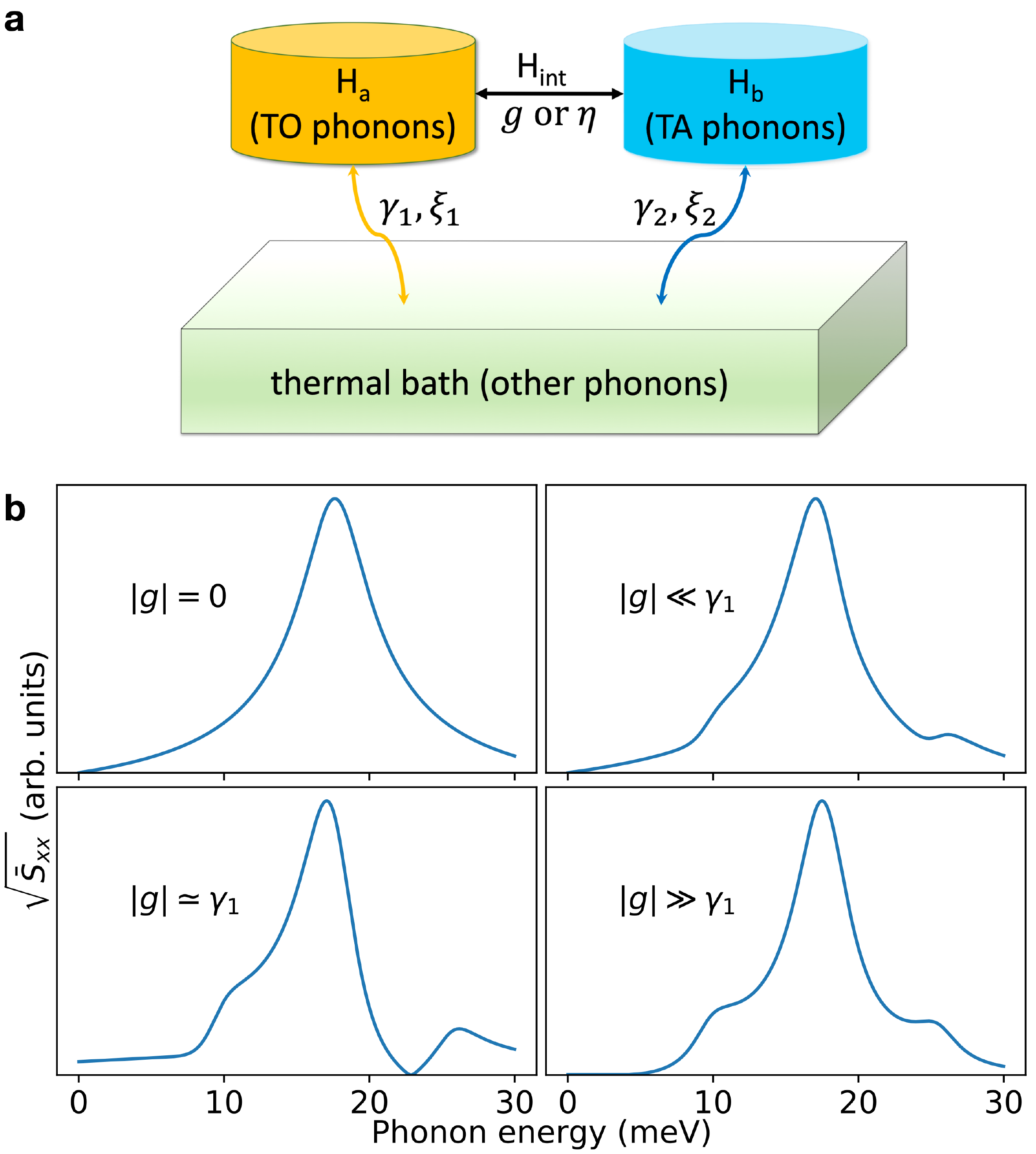}
    \caption{\textbf{Phonon self-transduction block diagram and its features.} \textbf{a,} The TO phonons and the TA phonons within 7-9\,meV are coupled by  phonon-phonon interactions. Meanwhile, they are in thermal equilibrium with the bath, which is an ensemble of other phonons.
    %with varying resonance frequencies. This Langevin model enable us to go beyond the perturbation framework of phonon study in solids. 
    \textbf{b,} %Power spectral density for coupling strengths in four coupling domains. 
    Power spectral density for coupling strength $\vert g \vert$ from none, weak, intermediate, strong.}
    \label{fig: diagram} 
\end{figure}

The general method of input-output theory \cite{Gardiner1985, Clerk2010} gives the Heisenberg-Langevin equations of motion for the two modes,
\begin{align}
    \dot{\hat{a}} &= -\mathrm{i}\omega_1 \hat{a}-\mathrm{i}\eta\hat{a}\left(\hat{b}^\dagger+\hat{b}\right)-\frac{\gamma_1}{2}\hat{a}-\sqrt{\gamma_1}\hat{\xi_1} \; , \label{HLeq_a} \\
    \dot{\hat{b}} &= -\mathrm{i}\omega_2 \hat{b}-\mathrm{i}\frac{\eta}{2}\left(\hat{a}^{\dagger}\hat{a}+ \hat{a}\hat{a}^{\dagger}\right)-\frac{\gamma_2}{2}\hat{b}-\sqrt{\gamma_2}\hat{\xi}_2 \; . \label{HLeq_b}
\end{align}
Here $\gamma_{1}$ and $\gamma_{2}$  are decay rates of the two modes, giving  phonon linewidths in energy. The other phonons are modeled as a thermal bath, described by  stochastic operators $\xi_{1}(t)$ and $\xi_{2}(t)$. These satisfy the correlation conditions: $\langle \hat{\xi}^{\dagger}(t)\hat{\xi}(t^{\prime})\rangle=n\delta(t-t^{\prime})$ and $\langle\hat{\xi}(t)\hat{\xi}^{\dagger}(t^{\prime})\rangle=(n+1)\delta(t-t^{\prime})$, where $n$ is the equilibrium Planck occupancy  $n=[\exp(\hbar\omega/k_{\rm B}T)-1]^{-1}$. 
%This is different from pure optical or optomechanical system, which will bring new physics features into the solid system.
These correlation conditions apply to both  modes 1 and 2; 
a situation that differs from optomechanical systems where correlations of
the stochastic variable $\xi_1$ for input noise of the optical photon
do not scale with equilibrium thermal occupancies.
Figure~\ref{fig: diagram}a depicts relationships  between the TA and TO phonons and the thermal bath of other phonons, showing   correspondences to th physical quantities of input-output theory. 

Using the concept of intermodulation, 
%where the vibrational spectral weight of TO modes is assigned to a central frequency $\omega_1$ with sidebands, 
a classical analysis by representing the phonon amplitudes of $\hat a$ as Fourier decomposition of sidebands \cite{Chan2011} shows that $\hat{a}$ comprises the frequency components  $\omega_1$ (first-order), $\omega_1 \pm \omega_2$ (second-order distortion), $\omega_{1}\pm 2\omega_{2}$ (third-order distortion), etc. 
%As we go beyond the usual linearization around a classical large amplitude of $\hat{a}$, we will explore 
Second-order effects are identified by transforming to 
a frame moving at the central frequency  $\omega_1$  
by replacing $\hat{a}(t)\rightarrow [\alpha+\hat{c}(t)]e^{-\mathrm{i}\omega_1t}$ and $\hat{\xi}_1(t)\rightarrow [\xi_{\rm in}+\hat{\xi}_1(t)] e^{-\mathrm{i}\omega_1 t}$,
where we take $\alpha$ to be real without loss of generality. 
This gives linearized equations of motion 
\begin{align}
    \label{eq: Langevin}
    \dot{\hat{c}} &= -\mathrm{i}g\left(\hat{b}^\dagger+\hat{b}\right)-\frac{\gamma_1}{2}\hat{c}-\sqrt{\gamma_1}\hat{\xi_1}\;,\\
    \dot{\hat{b}} &= -\mathrm{i}\omega_2 \hat{b}-\mathrm{i}g\left(\hat{c}^\dagger+\hat{c}\right)-\frac{\gamma_2}{2}\hat{b}-\sqrt{\gamma_2}\hat{\xi}_2\;,
\end{align}
where $g=\eta\alpha$ is the coupling strength. A straightforward calculation (see Supplemental) obtained the symmetrized power spectral density of displacement as
\begin{align}
    \bar{S}_{xx}[\omega] = &\frac{\hbar\gamma_1 \left(n_1+\frac{1}{2}\right)}{2m\omega_1} \Big(\left|\chi_{a,-}+2\mathrm{i}\omega_2g^2\chi_{a,-}^2\chi_{b,-}\bar{\chi}_{b,-}\right|^2  \nonumber \\
    &+\left|\chi_{a,+}-2\mathrm{i}\omega_2g^2\chi_{a,+}^2\chi_{b,+}\bar{\chi}_{b,+}\right|^2\Big) \; , \label{fullspectraldensity}
\end{align}
where the response functions are defined as
\begin{align}
    \chi_{a,\pm}^{-1} &=  -\mathrm{i}(\omega \pm \omega_1)+\frac{\gamma_1}{2} \; , \\
    \chi_{b,\pm}^{-1} &= -\mathrm{i}(\omega \pm \omega_1 - \omega_2)+\frac{\gamma_2}{2} \; ,\\
    \bar{\chi}_{b,\pm}^{-1} &= -\mathrm{i}(\omega \pm \omega_1 + \omega_2)+\frac{\gamma_2}{2} \; .
\end{align}

The first term in parens in Eq. \ref{fullspectraldensity}, 
\begin{align}
    \bar{S}_{xx}^{(+)}[\omega]=\frac{\hbar\gamma_1\left(n_1+\frac{1}{2}\right)}{2m\omega_1}\left|\chi_{a,-}+2\mathrm{i}\omega_2g^2\chi_{a,-}^2\chi_{b,-}\bar{\chi}_{b,-}\right|^2 \; ,
    \label{ghostmodeweight}
\end{align}
contributes spectral weight primarily to the positive frequency region. 
The other term $\bar{S}_{xx}^{(-)}$ contributes
to the negative.

In the absence of phonon coupling, i.e., $g=0$, 
Eq. \ref{ghostmodeweight} reduces to the thermal noise spectrum of a damped harmonic oscillator 
\begin{align}
    \bar{S}_{xx}^{(+),\mathrm{th}}[\omega]=\frac{\hbar\gamma_1\left(n_1+\frac{1}{2}\right)}{2m\omega_1}\frac{1}{(\omega-\omega_1)^2+(\gamma_1/2)^2} \; . \label{ReducedToDampedOscillator}
\end{align}
Figure~\ref{fig: diagram}b shows this is a Lorentzian function centered at $\omega_1$. Three other  cases are shown: weak coupling ($|g| \ll \gamma_1$), medium coupling ($|g| \simeq \gamma_1$), strong coupling ($|g| \gg \gamma_1$). 
To identify an IPS in a real material, the phonon-phonon interactions must be at least in the medium coupling regime. 
Recently, we identified NaBr with the rocksalt structure
as a highly anharmonic solid system \cite{Shen2020}, 
so it seemed an appropriate candidate for finding 
phonon intermodulation phenomena. 

%With work on NaBr, we  report the experimental evidence for new spectral features,
%compare it with the predictions from the ab initio computations, and show how our new phonon intermodulation model  explains these new features in the phonon spectrum of NaBr.

\section{Results}

\begin{figure*}[!htb]
    \centering
    \includegraphics[width=\linewidth]{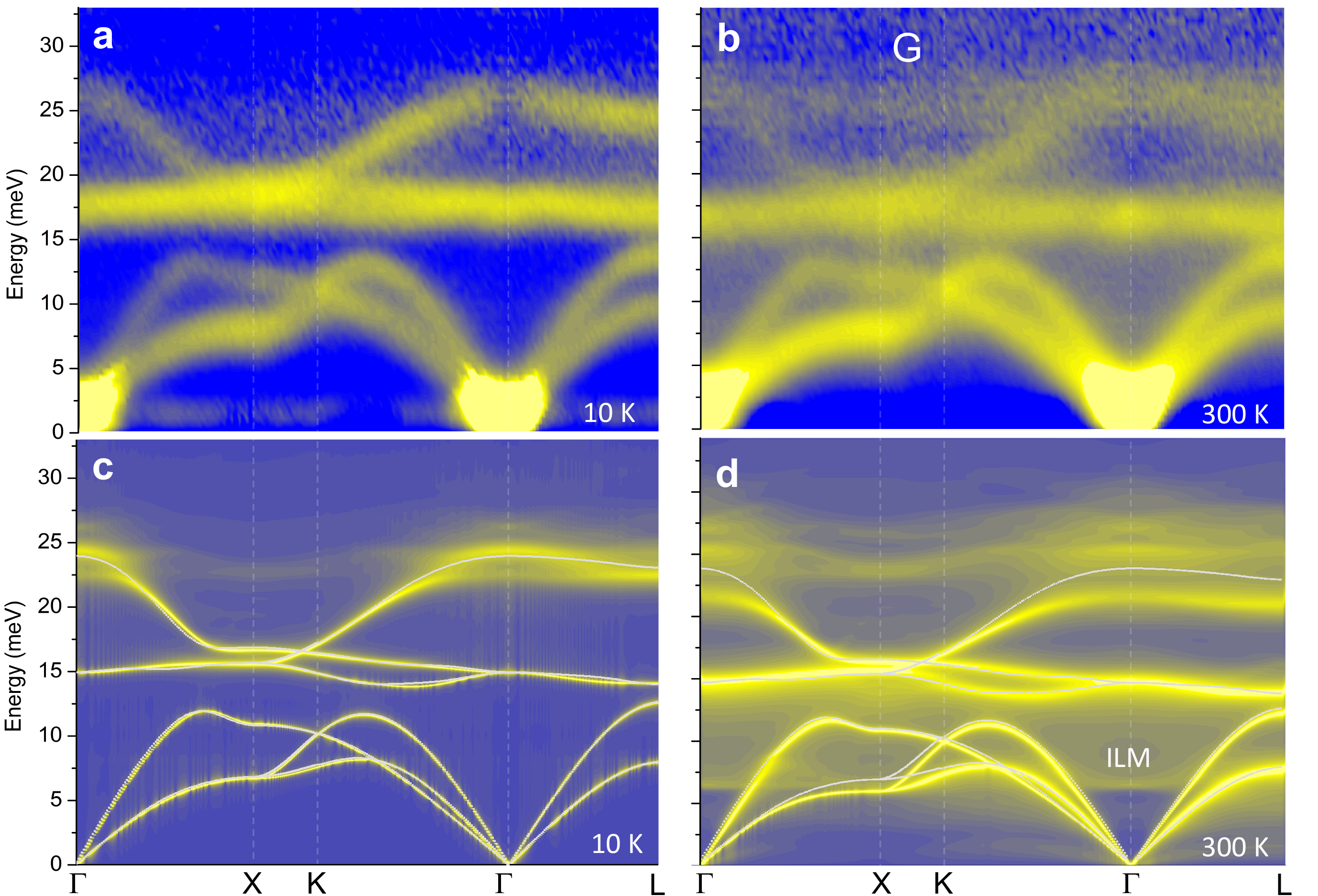}  %{exp_vs_cal.png}
    \caption{\textbf{Comparison between experimental and computational phonon dispersions of NaBr.} \textbf{a-b,} 2D slices through the four-dimensional scattering function  $S(\mathbf{Q}, \varepsilon)$, where $\varepsilon = \hbar \omega$, along  high symmetry lines in the first Brillouin zone. a,b are linear plots, with intensities corrected for thermal populations. \textbf{c-d,} Phonons in NaBr calculated with the quasiharmonic approximation (thin white  lines) and the full phonon spectral function  with phonon self-energy corrections. Temperatures are labeled. The intermodulation phonon sideband (IPS) ``G'' is seen in the experimental and computational results around the X point at 300\,K. The calculation also shows an ILM near the $\Gamma$-point at 300\,K. c,d are logarithmic plots of spectral weights. 
    %, but that is not found by measurements. 'G' refers to the ghost phonon branch. 
    }
    \label{fig: exp_vs_cal} 
\end{figure*}

A new spectral feature, labeled ``G" (ghost)
%{\color{red} ``USB''} 
in Fig.~\ref{fig: exp_vs_cal}b, appears at 300\,K. It is  flat over the Brillouin zone with an energy of 25-26\,meV. This new feature does not belong to any of the six  phonon branches expected for
the rocksalt structure (as in the white dotted lines 
in Figs. \ref{fig: exp_vs_cal}c,d
from the quasiharmonic approximation. %described in the Supplemental). 

\begin{figure*}[!htb]
    \centering
    \includegraphics[width=\linewidth]{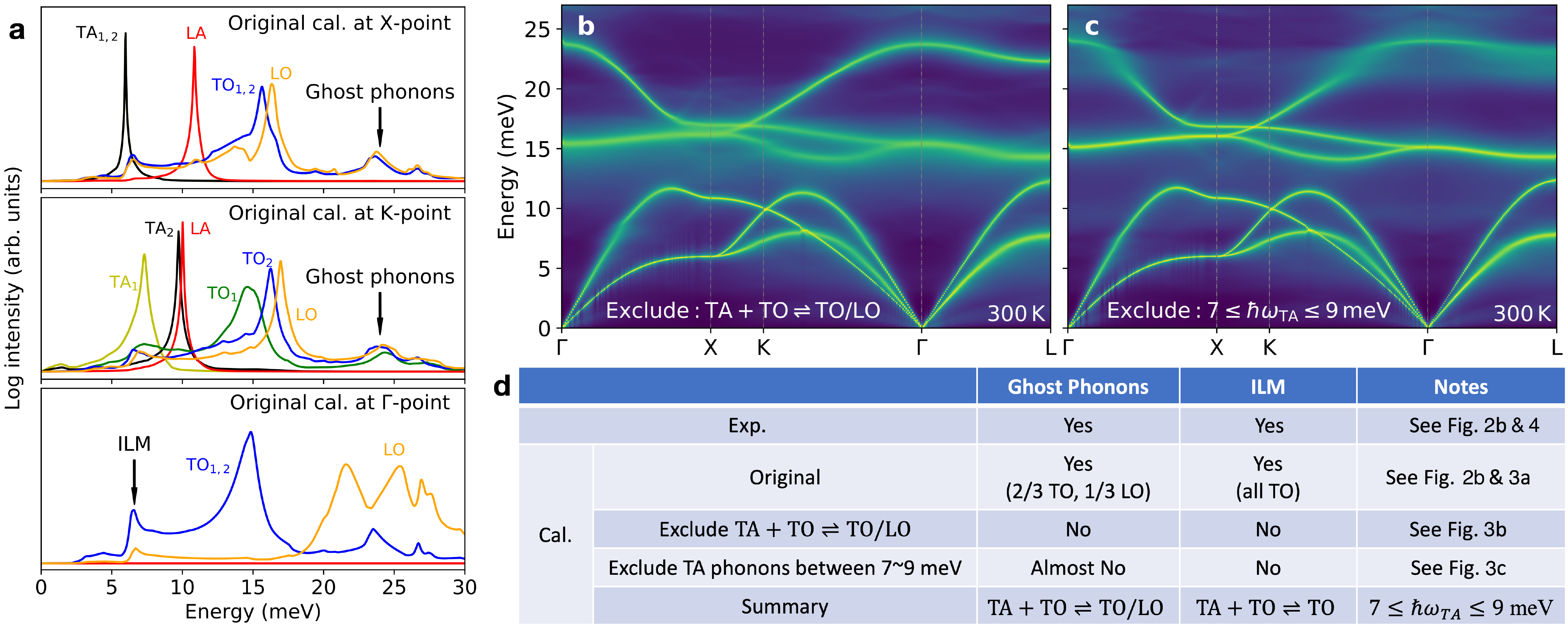}
    \caption{\textbf{Three-phonon processes associated with the IPS and ILM.} \textbf{a,} Calculated phonon lineshapes at the high symmetry points of $X$, $K$ and $\Gamma$. The first two were used to identify the components of the IPS, and the calculated ILM is shown in the bottom panel.  The phonon spectral function was recalculated \textbf{b,} without the three-phonon processes of $\mathrm{TA}+\mathrm{TO}\rightleftharpoons\mathrm{TO/LO}$, and \textbf{c,} without TA phonons between 7-9\,meV included in the three-phonon processes, compared with the main result in Fig.~\ref{fig: exp_vs_cal}d. \textbf{d,} Table of phonon processes for IPS and the ILM.}
    \label{fig: source} 
\end{figure*}

Figures~\ref{fig: exp_vs_cal}a,c show that at 10\,K, the  quasiharmonic and anharmonic calculations agree well with each other, and with the experimental phonon dispersions.
At 300\,K, the quasiharmonic model 
%and the calculations without the
%phonon self-energy corrections from Eq. \ref{eq: CubicHamiltonian}
predicts neither the phonon broadening, nor most of the thermal energy shift. 
The anharmonic calculations, however, reproduce these features and further predict the ghost intensity  around 25\,meV, 
in good agreement with experiment. % (labeled ``G'').
Figure~\ref{fig: source} shows that the ghost disappears when the calculation neglects the three-phonon anharmonic interactions of $\mathrm{TA}+\mathrm{TO}\rightleftharpoons\mathrm{TO/LO}$. 
Moreover, the participating TA phonons were shown to have energies between 7 and 9\,meV. 
%Therefore, the ghost phonon branch is generated through phonon-phonon interactions of $\mathrm{TA}+\mathrm{TO}\rightleftharpoons\mathrm{TO/LO}$ where the TA phonons are within 7-9\,meV. 
The diffuse features nearly vanish at 10\,K. 
(They should not vanish entirely, however, owing to 
effects of the zero-point occupancies of the TA and TO modes \cite{Markland2018, Kim2018}.)
Finally, the computations showed that the new features comprise  optical modes, with polarizations  distributed evenly over
all transverse (two) and longitudinal (one) possibilities. 

\begin{figure}
    \centering
    \includegraphics[width=\linewidth]{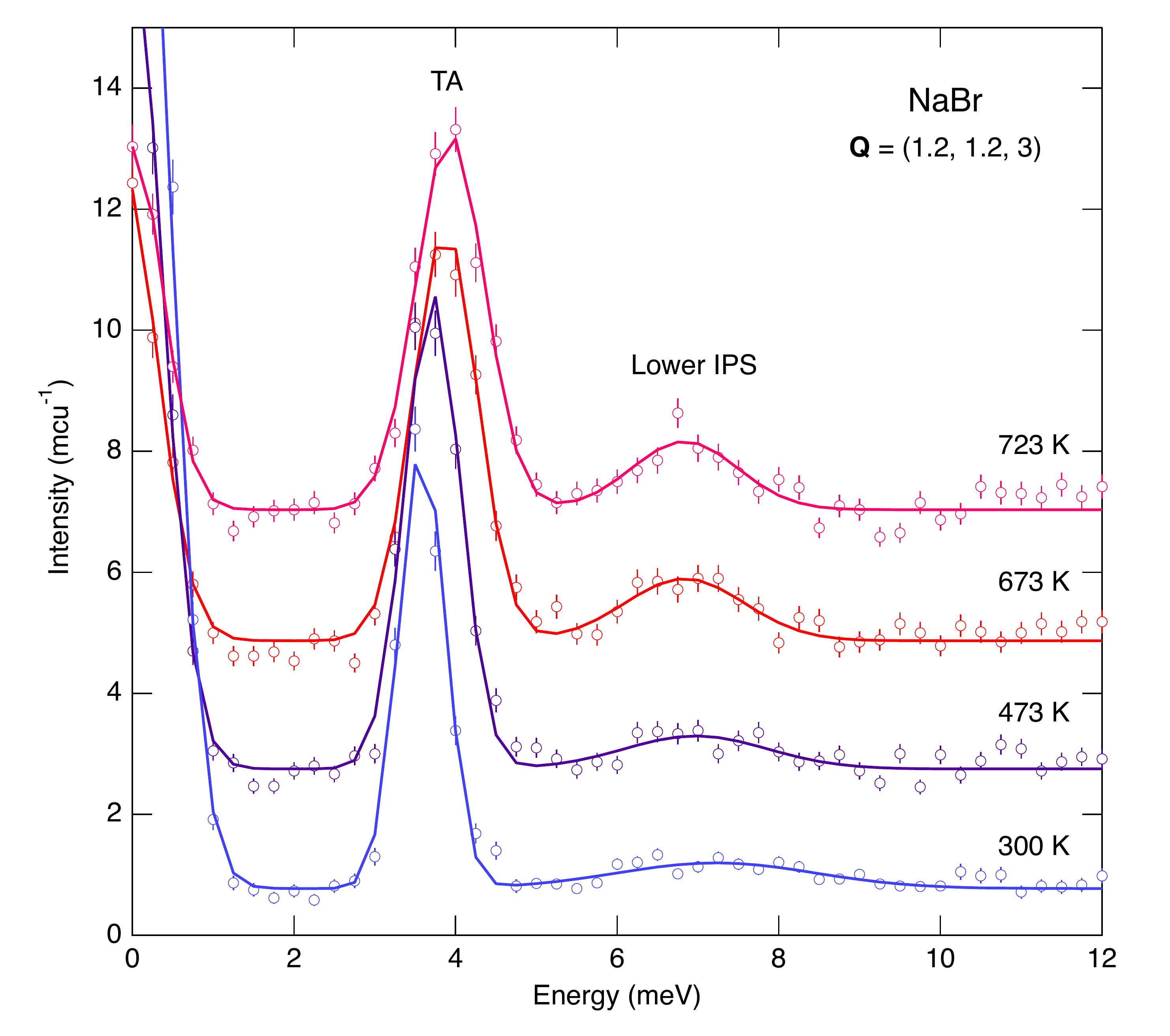}
    \caption{\textbf{Triple-axis energy scan}, showing the
    temperature dependence of the spectral intensity distribution for the TA and lower IPS phonons at $\vec{Q} = (1.2, 1.2, 3)$. 
    }
    \label{fig: tripleaxis} 
\end{figure}

Similarly, the calculated ILM near the $\Gamma$-point
 is produced by $\mathrm{TA}+\mathrm{TO}\rightleftharpoons\mathrm{TO}$ in the calculation.
Although the ILM is 
not definitive in Fig. \ref{fig: exp_vs_cal}b at 300\,K, it is  well-resolved at higher
temperatures in the HB3 data of 
Fig. \ref{fig: tripleaxis}.
This figure
shows the temperature dependence of the lower IPS (i.e, the ILM) as observed in the (113) Brillouin zone at (1.2, 1.2, 3.0), along with  the TA phonon. The spectral weight of this sideband gradually sharpens and intensifies with increasing temperature, and it
shifts slightly to lower energy. 
%, as expected from the thermal softening of its two intermodulating phonons. 
The TA mode has an apparent stiffening with temperature, but this is an artifact from thermal expansion \footnote{
The HB3 spectrometer was operated to maintain a constant $\vec{Q}$ for 300\,K, 
without correcting for thermal expansion, which shrinks 
the Brillouin zones of the crystal. 
The measured $\vec{Q}$ therefore increases as the crystal expands,
and the shift of the TA mode in Fig.~\ref{fig: tripleaxis} is consistent with the slope of the TA dispersion.}.
The TA mode also broadens with increasing temperature, as expected from the stronger coupling strength with increasing temperature,  discussed below.
Finally, the LA mode is suppressed in the spectra of Fig.~\ref{fig: tripleaxis} because  $\vec{Q}$ is nearly perpendicular to the  polarization vector $\vec{e}$ of the LA mode, i.e., direction of atom displacements in the mode. (If the LA mode were visible, its temperature dependence  would follow approximately the TA mode.)

%However, the challenge remains as the phonon self-energy correction does not normally form any new phonon branch, since its probing energy is based on the eigenvalues of the existing dynamical matrix. It is well-known that for a primitive cell with $N$ atoms, there are $3N$ different phonon modes for the single crystal, i.e., there are only $3N$ eigenvalues. 

\begin{figure*}
    \centering
    \includegraphics[width=\linewidth]{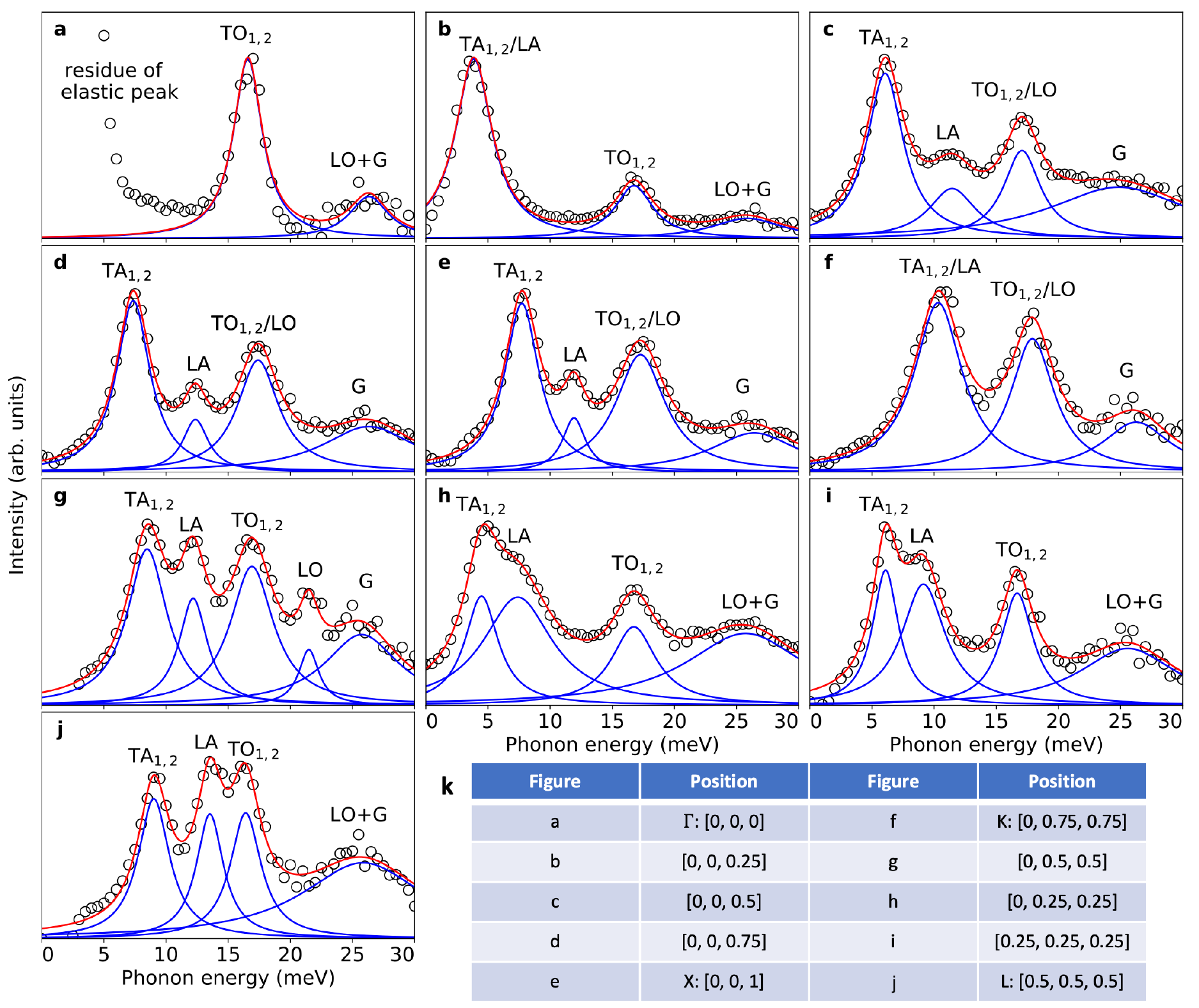}
    \caption{\textbf{Energy cuts at constant  $\mathbf{q}$ through experimental dispersions of Fig. \ref{fig: exp_vs_cal}b.} \textbf{a-j,} Experimental  data are points; fitted peaks are in blue, and the cumulative fitting results are in red. \textbf{k,} Table of the $\mathbf{q}$-points for each panel.}
    \label{fig: peaks} 
\end{figure*}

The three phonon modes in Eq. \ref{eq: CubicSecondOrder}
are eigenstates of a dynamical matrix. 
Small anharmonic shifts and broadenings of these 
eigenstates  do not produce new phonon branches or spectral features.
Our anharmonic calculations 
obtained the diffuse features after applying 
a Kramers-Kronig transformation of Eq. \ref{eq: Kramers-Kronig}
to the imaginary part of the phonon self energy
of Eq.~\ref{eq: CubicSecondOrder}. 
The semiquantitative success is interesting, 
because the calculations also predict a weak ILM \cite{Manley2019, Shulumba2017}.
The dispersions in Fig. \ref{fig: exp_vs_cal}c,d
are on a logarithmic scale, however, and 
the ILM and ghost modes are 
much weaker in the calculation with perturbation theory
than in the experimental intensities,
discussed with Fig. \ref{fig: peaks}.

Compared to Eq.~\ref{eq: CubicSecondOrder},
the Heisenberg--Langevin 
Eqs. \ref{HLeq_a} and \ref{HLeq_b}
have no implicit assumption that $\mathcal{H}_3$ is small. 
The intensities of the measured
ghost modes are seen in Fig. \ref{fig: peaks},
which are 
energy cuts at different $Q$ along
high-symmetry directions  
through the experimental
data of Fig. \ref{fig: exp_vs_cal}b. 
The points near $X$ or $K$ (see Fig.~\ref{fig: peaks}c-g), show an extra peak above 20\,meV, which is distinct from the highest normal LO phonon branch.

Phonon centroids were obtained
by fitting with the Levenberg--Marquardt nonlinear least square method for multiple Lorentzian functions, giving the fitting parameters listed in Table S1 in the Supplementary Information. 
The fitting results were used to obtain average   energies 
and linewidths of 
TO phonons and  TA phonons in the energy range of  7-9\,meV (i.e. $\omega_1$, $\omega_2$, $\gamma_1$, $\gamma_2$) for calculating sidebands with the Heisenberg--Langevin model.
By performing averages over TA and TO peaks that
were not impaired by overlaps with other peaks, 
$\omega_{1} = 16.97(10)$,
$\omega_{2} =  8.2(2)$,
$\gamma_{1} = 3.6(10) $, 
$\gamma_{2} =  3.7(7)$\,meV,
where subscripts 1 and 2 denote TO and TA, respectively
The average frequency of the ghost mode ($\omega_{\rm G}$) was obtained 
from Fig. \ref{fig: peaks} panels c,d,e,f,g as
$\omega_{\rm G} = 25.9(5)$\,meV, 
satisfying $\omega_1 + \omega_2 \simeq \omega_{\rm G}$.

The coupling strength was calculated by comparing the power intensity between the
Heisenberg--Langevin model and the
measured peak intensities $I_{\rm exp}(\omega)$ 
at 300\,K. Their ratio avoids scaling factors
\begin{eqnarray}
    \frac{\bar{S}_{xx}^{(+)}[\omega_1]}{\bar{S}_{xx}^{(+)}[\omega_1+\omega_2]} =
     \frac{I_{\rm exp} (\omega_1) \ n(\omega_1)}{I_{\rm exp} ( \omega_1+\omega_2) \ n(\omega_1+\omega_2) \ 2/3 } \; .
     \label{SidebandIntensityRatio}
\end{eqnarray}
%\begin{align}
%    \frac{\bar{S}_{xx}^{(+)}[\omega_1]}{\bar{S}_{xx}^{(+)}[\omega_1+\omega_2]} = \frac{(\mathrm{Measured\ Phonon\ Intensity\ Peak\ at\ }\omega = \omega_1)*n(\omega_1)}{(\mathrm{Measured\ Phonon\  Intensity\ Peak\ at\ }\omega = \omega_1+\omega_2)*n(\omega_1+\omega_2)*2/3 } \; .
%\end{align}
The terms in the left-hand side are derived in the Supplemental Information. 
Here $n(\omega)=[\exp(\hbar\omega/k_{\rm B}T)-1]^{-1}$ is the Planck distribution function (the thermal weight was corrected for the measured intensity shown in Fig. \ref{fig: exp_vs_cal}a,b), and 
the $2/3$ factor is included because two-thirds of the IPS feature are G-TO phonons.
%The physics solution corresponds to the one with in the Supplemental with
%an observable upper-shifted sideband (AS peak).

\section{Discussion \label{Discussion}}

At room temperature, our analysis showed that the lower intermodulation phonon sideband (IPS) should not be visible as a distinct peak. Also, the ARCS spectrometer  has lower energy resolution at the energy of the lower sideband than  at the energy of the upper sideband. The lower sideband is  better seen with 
the HB3 instrument at higher temperatures (Fig. \ref{fig: tripleaxis}). 

%with the same visibility as the ghost phonon mode. 
%The Supplemental shows the ILM with more focused
%triple-axis measurements, but its lower visibility seems beyond our calculations based on the perturbation theory.
%challenge also remains due to the ``missing'' ILM predicted by the calculation, as the same computational and experimental techniques have been successful on the ILM in PbSe 

%In what follows we present an alternative physical explanation for  the phonon dispersions of NaBr, inspired by interpretations of  modern laser-cavity experiments \cite{Clerk2010, Kippenberg2008, Vahala2009, Chan2011, Benz2016, Riedinger2016, Renninger2018, Tavernarakis2018}. 
%Our new model follows Eq.~\ref{eq: CubicHamiltonian}, with $j$, $j^{\prime}$ and $j^{\prime\prime}$ designating the TA , TO and diffuse  phonon features, respectively, but there is 

The $\mathbf{q}$-dependence of phonons in solids is not
considered in Eqs. \ref{HLeq_a} and \ref{HLeq_b}, 
and the  conservation 
of crystal momentum 
%in Eq.~\ref{eq: CubicHamiltonian} 
is  an added complexity that is not needed for other coupled quantum systems \cite{Clerk2010, Kippenberg2008, Vahala2009, Chan2011, Benz2016, Riedinger2016, Renninger2018, Tavernarakis2018}. Figure~\ref{fig: exp_vs_cal} shows that in NaBr, however, the  TO phonon branch and the 
upper IPS are largely flat and dispersionless.
The TA phonons involved in the three-phonon processes are in a small energy range of 7-9\,meV as shown above with Fig. \ref{fig: source}. 
The phonon dispersions (Fig.~\ref{fig: exp_vs_cal})
show that most of the TA phonons are in this energy range, 
forming plateaus reaching to the Brillouin zone boundary. 
Most TA phonons can be described with 
an average energy independent of  $\mathbf{q}$,
and hence a coupling strength parameter independent of $\mathbf{q}$.

The conservation of crystal momentum, $\mathbf{q}+\mathbf{q}^{\prime} = \mathbf{q}^{\prime\prime} + k\mathbf{G}$ ($k=0,1$) allows the $\mathbf{q}^{\prime\prime}$ of the diffuse modes to sweep over all the first Brillouin zone when $\mathbf{q}^{\prime}$ (TO modes)  covers the first Brillouin zone, even for a single value of $\mathbf{q}$. 
Finally, the anharmonicity in  Na-Br 
has no strong dependence on crystallographic
direction, and
is dominated by 
first-neighbor interactions \cite{Shen2020}.

Consider first the case where {TA + TO $\rightleftharpoons$ G-LO} and the interacting TA, TO, and LO diffuse phonon modes can be treated as individual quantum oscillators with a coupling coefficient $\eta$. The total Hamiltonian is the same as Eq.~\ref{eq: HamiltonianForLOGhostPhononsSimplified}. After dropping terms that do not conserve energy,
\begin{align}
    \mathcal{H_{\rm sys}}\,=\ \mathcal{H}_0+\hbar\eta\left(\hat{a}_{j}\hat{a}_{j^\prime}   \hat{a}^\dagger_{j^{\prime\prime}} + \hat{a}^\dagger_{j} \hat{a}^\dagger_{j^\prime} \hat{a}_{j^{\prime\prime}} \right) \;.
\end{align}
This is the same form as for parametric down-conversion in nonlinear optics. The mode coupling is enhanced resonantly when $\omega_{j^{\prime}}=\omega_{j^{\prime\prime}}-\omega_{j}$.

The second case {TA + TO $\rightleftharpoons$ G-TO} has the same transverse polarization for two optical modes. These two TO modes can be modeled as a single oscillator. Its  spectral weight is re-distributed in energy owing to strong coupling to the TA mode. This is exactly the phonon intermodulation mechanism described with the quantum
Langevin model.

The Supplemental Information gives more
details of how experimentally 
measured parameters of $\omega_1\simeq 16.97\,\mathrm{meV}$, $\omega_2\simeq 8.2\,\mathrm{meV}$, $\gamma_1\simeq 3.6\,\mathrm{meV}$ and $\gamma_2\simeq 3.7\,\mathrm{meV}$
were used in a numerical analysis that generated the 
spectral shapes 
of Fig.~\ref{fig: diagram}b, with different coupling
parameters, $g$. 
In the weak-coupling case, 
our measured INS spectra would  show no features
other than the main peak at $\omega=\omega_1$. 
In the medium-coupling case, the lower sideband 
peak at $\omega_1-\omega_2$ 
is only a shoulder on the main peak, but the
upper sideband with $\omega_1+\omega_2$ should be
distinct. 
%This asymmetry in the spectral sidebands is similar to the optomechanical cooling mechanism. 
The strong-coupling case shows two symmetric sidebands
as shoulders on the main peak.
The clear, isolated diffuse intensity ``G''  in 
Fig. \ref{fig: exp_vs_cal}b 
and weaker ILM are consistent with the TA and TO modes being in the medium-coupling case.
To verify this, we solved numerically  for the ratio between the heights of the two resonant peaks at $\omega=\omega_1$ and $\omega=\omega_1+\omega_2$ (Eq. \ref{SidebandIntensityRatio}), to obtain the coupling strength parameter $|g|\simeq 3.7\,\mathrm{meV}$, showing that the system is indeed in the medium-coupling domain. This also explains the difference in visibility between the ILM and the ghost phonon mode.
%It also shows the limitation of the lattice dynamics calculations based on the perturbation theory, as part of its prediction on the Stoke peak is no longer valid in the medium-coupling domain, and thus the calculated possible ILM was not detected by the same measurement. This also indicates a new way to solve the puzzle of ILMs by thinking outside the perturbation box. 

The prior treatment of ILMs \cite{Sievers1988} considered the dynamics of a classical system with linear and cubic terms in the restoring forces between neighboring atoms, and showed the conditions for mode localization. Our approach with the quantum Langevin equation is better able to predict the spectral shape, providing deeper insights into the phonon intermodulation mechanism.
Unlike a classical intermodulation, phonon intermodulation can have an asymmetric quantum effect of enhancing one sideband at the expense of  the other. 

Other materials with anharmonic phonons should 
have IPSs at modest temperatures. 
Different alkali halides are obvious candidates, as are  materials 
with phonon instabilities, 
where nonlinear phonon interactions
may generate sidebands as the instabilities grow. 
The sidebands in NaBr were  from acoustic plus optic modes,
but in principle, two anharmonic optic modes could also
generate sidebands.
Anharmonicity may offer a new functionality for optical materials
in the infrared, or a means to modulate visible light in ways 
that originate with phonon interactions,
rather than an asymmetry in electronic polarizability. 
The ghost modes in NaBr  decay rapidly  into TA and TO modes
through three-phonon processes. 
The two new phonons are in phase, and would be entangled in the
Einstein-Podolsky-Rosen sense.  
Their coherence time will be short, however.

%Excitations in the IPSs are mixtures of TA and TO phonons, entangled in the EPR sense. 
%Pumping the upper IPS at low temperature could create entangled phonons of controlled polarization, which may be suitable for probing fast processes in crystals. 
%The coherency of these entangled states 
%is not high, however, and testing this seems difficult. 

\section{Conclusion}

A new band of spectral
intensity from high energy phonons is predicted and
observed in NaBr.
It is an intermodulation phonon sideband (IPS)
from anharmonic interactions between  normal modes. 
%In conclusion, we found a new diffuse phonon intensity of mixed polarization
%in NaBr at 300\,K. 
Its partner, the lower sideband, is an intrinsic localized mode (ILM).
The transfer of spectral weight to upper and lower IPSs likely
occurs in other anharmonic materials, but the flat dispersions in
NaBr make them easier to observe. 
The TO part of this feature 
is consistent with an IPS from the anharmonic coupling
of TO  modes and  TA modes. 
The LO part is consistent with strong 
three-phonon process, again with anharmonic coupling to the TA modes. 
The spectral shapes and weights of the IPSs are altered by the quantum back action from the thermal bath. There are similarities to the formation of sidebands in laser-cavity experiments, which also depend on anharmonicity and quantum force fluctuations from the thermal bath. 
Compared to laser-cavity experiments with photons, 
the anharmonic sidebands in NaBr 
are a natural process that occurs in 
thermodynamic equilibrium, 
and both the interacting modes have the
noise spectrum from the thermal bath.
The IPS should be present at 0\,K owing to couplings to the 
zero-point levels, and some traces may be visible in the dispersions at 10\,K. 
The spectral shapes of the
two sidebands offer a probe of  quantum noise, 
giving parameters for mode coupling, and damping from the thermal bath. 
Perhaps the upper IPS could offer new methods for the thermal control of light-matter interactions.

%\bibliographystyle{apsrev4-1}
%\bibliography{references}
% Produces the bibliography via BibTeX.
\providecommand{\noopsort}[1]{}\providecommand{\singleletter}[1]{#1}%

\section{Acknowledgements}

We thank O. Hellman, K. Vahala and F. Yang for helpful discussions. Research at the Spallation Neutron Source (SNS) and the High Flux Isotope Reactor (HFIR) at the Oak Ridge National Laboratory was sponsored by the Scientific User Facilities Division, Basic Energy Sciences (BES), Department of Energy (DOE). M.E.M was supported by the US Department of Energy, Office of Science, Office of Basic Energy Sciences, Materials Sciences and Engineering Division under Contract Number DE-AC05-00OR22725. This work used resources from National Energy Research Scientific Computing Center (NERSC), a DOE Office of Science User Facility supported by the Office of Science of the US Department of Energy under Contract DE-AC02-05CH11231. This work was supported by the DOE Office of Science, BES, under Contract DE-FG02-03ER46055.

\end{document}

\title{Supplementary Information for:\\ ``Prediction and Observation of Intermodulation Sidebands from Anharmonic Phonons in NaBr''}

\author{Y. Shen}
\email{yshen@caltech.edu}
\affiliation{Department of Applied Physics and Materials Science, California Institute of Technology, Pasadena, California 91125, USA}
\author{C. N. Saunders}
\affiliation{Department of Applied Physics and Materials Science, California Institute of Technology, Pasadena, California 91125, USA}
\author{C. M. Bernal}
\affiliation{Department of Applied Physics and Materials Science, California Institute of Technology, Pasadena, California 91125, USA}
\author{D. L. Abernathy}
\affiliation{Neutron Scattering Division, Oak Ridge National Laboratory, Oak Ridge, Tennessee 37831, USA}
\author{T. J. Williams}
\affiliation{Neutron Scattering Division, Oak Ridge National Laboratory, Oak Ridge, Tennessee 37831, USA}
\author{M. E. Manley}
\affiliation{Material Science and Technology Division, Oak Ridge National Laboratory, Oak Ridge, Tennessee 37831, USA}
\author{B. Fultz}
\email{btf@caltech.edu}
\affiliation{Department of Applied Physics and Materials Science, California Institute of Technology, Pasadena, California 91125, USA}

\date{\today}% It is always \today, today,

\maketitle

\section{Classical Analysis}

Consider the original Heisenberg-Langevin equations (Eq. 6-7 in the main text):
\begin{align}
    \label{eq: OriHL}
    \dot{\hat{a}} &= -\mathrm{i}\omega_1 \hat{a}-\mathrm{i}\eta\hat{a}\left(\hat{b}^\dagger+\hat{b}\right)-\frac{\gamma_1}{2}\hat{a}-\sqrt{\gamma_1}\hat{\xi_1} \; , \\
    \dot{\hat{b}} &= -\mathrm{i}\omega_2 \hat{b}-\mathrm{i}\frac{\eta}{2}\left(\hat{a}^{\dagger}\hat{a}+ \hat{a}\hat{a}^{\dagger}\right)-\frac{\gamma_2}{2}\hat{b}-\sqrt{\gamma_2}\hat{\xi}_2 \; .
\end{align}
Following the method in Ref.~\cite{Chan2011}, we can replace the phonon amplitudes with a Fourier decomposition of sidebands as
\begin{align}
    \hat{a}\longrightarrow \alpha= \sum_{j}A_j e^{-\mathrm{i}\omega_{a, j}t},\qquad b\longrightarrow B_0e^{-\mathrm{i}\omega_2 t} \; ,
\end{align}
where $j$ is the sideband order.
Substituting these into Eq.~\ref{eq: OriHL}
\begin{align}
    \sum_{j}-\mathrm{i}\omega_{a, j}A_j e^{-\mathrm{i}\omega_{a, j}t} = -\left(\mathrm{i}\omega_1+\frac{\gamma_1}{2}\right) \sum_{j}A_j e^{-\mathrm{i}\omega_{a, j}t} -\mathrm{i}\eta B_0 \sum_{j}A_j \left(e^{-\mathrm{i}(\omega_{a, j}+\omega_2) t} + e^{-\mathrm{i}(\omega_{a, j}-\omega_2) t}\right) \; . \label{ClassicalModulation}
    %-\sqrt{\gamma_1}\hat{\xi_1}
\end{align}
Comparing the two sides of Eq. \ref{ClassicalModulation}, we find the  intermodulation frequencies
\begin{align}
    \mathrm{First}\ \mathrm{order:}\ &j = 1,\ \omega_{a, 1}=\omega_1; \nonumber \\
    \mathrm{Second}\ \mathrm{order:}\ &j = 2,\ \omega_{a, 2}=\omega_1\pm\omega_2; \nonumber \\
    \mathrm{Third}\ \mathrm{order:}\ &j = 3,\ \omega_{a, 3}=\omega_{a, 2}\pm\omega_2=\omega_1\pm2\omega_2; \nonumber \\
    \ldots
\end{align}

\section{Solving the quantum Langevin equations}
The Heisenberg-Langevin equations in the main text (Eq. 7) can be solved by Fourier transformation.For
an operator in the time domain $\hat{O}(t)$, we use an operator in the frequency domain, $\hat{O}[\omega]$
\begin{align}
    \hat{O}[\omega]= \frac{1}{\sqrt{2\pi}}\int_{-\infty}^{+\infty}\mathrm{d}t e^{\mathrm{i}\omega t}\hat{O}(t) \; , \\
    \hat{O}^{\dagger}[\omega]= \frac{1}{\sqrt{2\pi}}\int_{-\infty}^{+\infty}\mathrm{d}t e^{\mathrm{i}\omega t}\hat{O}^{\dagger}(t) \; .
\end{align}
This procedure gives the following Fourier transformed equations
\begin{align}
    (-\mathrm{i}\omega+\frac{\gamma_1}{2})\hat{c}[\omega] &= -\mathrm{i}g\left(\hat{b}^\dagger[\omega]+\hat{b}[\omega]\right)-\sqrt{\gamma_1}\hat{\xi_1}[\omega] \; , \\
    (-\mathrm{i}\omega+\frac{\gamma_1}{2})\hat{c}^\dagger[\omega] &= \mathrm{i}g\left(\hat{b}^\dagger[\omega]+\hat{b}[\omega]\right)-\sqrt{\gamma_1}\hat{\xi_1}^\dagger[\omega] \; ,  \\
    \left[-\mathrm{i}(\omega-\omega_2)+\frac{\gamma_2}{2}\right]\hat{b}[\omega] &= -\mathrm{i}g\left(\hat{c}^\dagger[\omega]+\hat{c}[\omega]\right)-\sqrt{\gamma_2}\hat{\xi}_2[\omega] \; , \\
    \left[-\mathrm{i}(\omega+\omega_2)+\frac{\gamma_2}{2}\right]\hat{b}^\dagger[\omega] &= \mathrm{i}g\left(\hat{c}^\dagger[\omega]+\hat{c}[\omega]\right)-\sqrt{\gamma_2}\hat{\xi}_2^\dagger[\omega] \; . 
\end{align}
Solutions forf $\hat{c}$ and $\hat{c}^{\dagger}$ are
\begin{align}
    \hat{c}[\omega]&=\chi_c^2\chi_b\bar{\chi}_b \left\{ -\sqrt{\gamma_1} \left[ \left( \chi_c^{-1}\chi_b^{-1} \bar{\chi}_b^{-1}+2\mathrm{i}\omega_2 g^2 \right)\hat{\xi}_1+2\mathrm{i}\omega_2 g^2\hat{\xi}_1^\dagger\right] +\mathrm{i}g\sqrt{\gamma_2}\chi_c^{-1}\left(\bar{\chi}_b^{-1}\hat{\xi}_2+\chi_b^{-1}\hat{\xi}_2^\dagger\right)\right\} \; , \\
    \hat{c}^\dagger[\omega] &= \chi_c^2\chi_b\bar{\chi}_b \left\{ -\sqrt{\gamma_1}\left[ -2\mathrm{i}\omega_2 g^2 \hat{\xi}_1+\left( \chi_c^{-1}\chi_b^{-1} \bar{\chi}_b^{-1}-2\mathrm{i}\omega_2 g^2 \right)\hat{\xi}_1^\dagger\right] -\mathrm{i}g\sqrt{\gamma_2}\chi_c^{-1}\left(\bar{\chi}_b^{-1}\hat{\xi}_2+\chi_b^{-1}\hat{\xi}_2^\dagger\right)\right\} \; ,
\end{align}
where the bare response functions are 
\begin{align}
    \chi_c^{-1}&=-\mathrm{i}\omega+\frac{\gamma_1}{2} \; , \\
    \chi_b^{-1}&=-\mathrm{i}(\omega-\omega_2)+\frac{\gamma_2}{2}  \; , \\
    \bar{\chi}_b^{-1}&=-\mathrm{i}(\omega+\omega_2)+\frac{\gamma_2}{2}  \; .
\end{align}

Transforming back to the original operators $\hat{a}$ ($\hat{a}^{\dagger}$), the position operator is obtained 
\begin{align}
    \hat{x} = x_{\rm zpf}\left(\hat{a}+\hat{a}^\dagger\right) =x_{\rm zpf}\left[\left(\alpha+\hat{c}\right)e^{-\mathrm{i}\omega_1t}+\left(\alpha+\hat{c}^\dagger\right)e^{\mathrm{i}\omega_1 t}\right] \; ,
\end{align}
with $x_{\rm zpf}=\sqrt{\frac{\hbar}{2m\omega_1}}$ as the amplitude of zero point fluctuations.
Correspondingly
\begin{align}
    x[\omega]=&\, x_{\rm zpf}\left(c[\omega-\omega_1]+c^{\dagger}[\omega+\omega_1]+ \alpha\delta[\omega-\omega_1]+\alpha\delta[\omega+\omega_1]\right)\; .
\end{align}
Neglecting the unimportant terms of single $\delta$-functions, the displacement power spectral density, $S_{xx}[\omega]$, can be obtained by
\begin{align}
    S_{xx}[\omega]&=\int_{-\infty}^{+\infty}\left\langle\hat{x}[\omega]\hat{x}[\omega^\prime]\right\rangle\mathrm{d}\omega^\prime \nonumber \\
    &=\frac{\hbar}{2m\omega_1}\int_{-\infty}^{+\infty} \left\langle \left(c[\omega-\omega_1]+c^{\dagger}[\omega+\omega_1]\right)\left(c[\omega^\prime-\omega_1]+c^{\dagger}[\omega^\prime+\omega_1]\right)\right\rangle \mathrm{d}\omega^\prime \nonumber \\
    &=\frac{\hbar}{2m\omega_1}\left( 
    \int_{-\infty}^{+\infty}
    \left\langle c[\omega-\omega_1]c[\omega^\prime-\omega_1]\right\rangle
    \mathrm{d}\omega^\prime + 
    \int_{-\infty}^{+\infty}\left\langle c[\omega-\omega_1]c^\dagger[\omega^\prime+\omega_1]\right\rangle
    \mathrm{d}\omega^\prime \right.\nonumber \\
    &\left. \quad\quad\quad\quad+ \int_{-\infty}^{+\infty}\left\langle c^\dagger[\omega+\omega_1]c[\omega^\prime-\omega_1]\right\rangle\mathrm{d}\omega^\prime +
    \int_{-\infty}^{+\infty}\left\langle c^\dagger[\omega+\omega_1]c^\dagger[\omega^\prime+\omega_1]\right\rangle
    \mathrm{d}\omega^\prime\right)
    \label{eq: Sxx}
\end{align}

In the frequency domain, the input noise operators satisfy the relations of 
\begin{align}
    \left\langle\hat{\xi}_{1,2}[\omega_1]\hat{\xi}_{1,2}^\dagger[\omega_2]\right\rangle&=(n_{1,2}+1)\delta[\omega_1+\omega_2] \; , \\
    \left\langle\hat{\xi}_{1,2}^\dagger[\omega_1]\hat{\xi}_{1,2}[\omega_2]\right\rangle&=n_{1,2}\delta[\omega_1+\omega_2] \; .
\end{align}
Employing these relations, every term in Eq.~\ref{eq: Sxx} can be calculated
%\begingroup
%\allowdisplaybreaks
\begin{align}
    \label{eq: example}
    &\int_{-\infty}^{+\infty}\left\langle c[\omega-\omega_1]c[\omega^\prime-\omega_1]\right\rangle\mathrm{d}\omega^\prime \nonumber \\
    =\,&\gamma_1\left|\chi_{a,-}^2\chi_{b,-}\bar{\chi}_{b,-}\right|^2\left(2\mathrm{i}\omega_2g^2\right) \left[\left(\chi_{a,-}^{-1}\chi_{b,-}^{-1}\bar{\chi}_{b,-}^{-1}+2\mathrm{i}\omega_2g^2\right)(n_1+1)+\left(\chi_{a,-}^{*-1}\chi_{b,-}^{*-1}\bar{\chi}_{b,-}^{*-1}+2\mathrm{i}\omega_2g^2\right)n_1\right]\nonumber \\
    &-g^2\gamma_2 \left|\chi_{a,-}\chi_{b,-}\bar{\chi}_{b,-}\right|^2\left(\left|\bar{\chi}_{b,-}^{-1}\right|^2(n_2+1)+\left|\chi_{b,-}^{-1}\right|^2n_2\right) \; ,\\
    &\int_{-\infty}^{+\infty}\left\langle c[\omega-\omega_1]c^\dagger[\omega^\prime+\omega_1]\right\rangle\mathrm{d}\omega^\prime \nonumber \\
    =\,&\gamma_1\left|\chi_{a,-}^2\chi_{b,-}\bar{\chi}_{b,-}\right|^2 \left[\left|\chi_{a,-}^{-1}\chi_{b,-}^{-1}\bar{\chi}_{b,-}^{-1}+2\mathrm{i}\omega_2g^2\right|^2(n_1+1)+4\omega_2^2 g^4n_1\right] \nonumber \\
    &+g^2\gamma_2 \left|\chi_{a,-}\chi_{b,-}\bar{\chi}_{b,-}\right|^2\left(\left|\bar{\chi}_{b,-}^{-1}\right|^2(n_2+1)+\left|\chi_{b,-}^{-1}\right|^2n_2\right) \; ,\\
    &\int_{-\infty}^{+\infty}\left\langle c^\dagger[\omega+\omega_1]c[\omega^\prime-\omega_1]\right\rangle\mathrm{d}\omega^\prime \nonumber \\
    =\,&\gamma_1\left|\chi_{a,+}^2\chi_{b,+}\bar{\chi}_{b,+}\right|^2 \left[4\omega_2^2 g^4(n_1+1)+\left|\chi_{a,+}^{-1}\chi_{b,+}^{-1}\bar{\chi}_{b,+}^{-1}-2\mathrm{i}\omega_2g^2\right|^2n_1\right] \nonumber \\
    &+g^2\gamma_2 \left|\chi_{a,+}\chi_{b,+}\bar{\chi}_{b,+}\right|^2\left(\left|\bar{\chi}_{b,+}^{-1}\right|^2(n_2+1)+\left|\chi_{b,+}^{-1}\right|^2n_2\right) \; ,\\
    &\int_{-\infty}^{+\infty}\left\langle c^\dagger[\omega+\omega_1]c^\dagger[\omega^\prime+\omega_1]\right\rangle\mathrm{d}\omega^\prime \nonumber \\
    =\,&\gamma_1\left|\chi_{a,+}^2\chi_{b,+}\bar{\chi}_{b,+}\right|^2\left(-2\mathrm{i}\omega_2g^2\right) \left[\left(\chi_{a,+}^{*-1}\chi_{b,+}^{*-1}\bar{\chi}_{b,+}^{*-1}-2\mathrm{i}\omega_2g^2\right)(n_1+1)+\left(\chi_{a,+}^{-1}\chi_{b,+}^{-1}\bar{\chi}_{b,+}^{-1}-2\mathrm{i}\omega_2g^2\right)n_1\right] \nonumber \\
    &-g^2\gamma_2 \left|\chi_{a,+}\chi_{b,+}\bar{\chi}_{b,+}\right|^2\left(\left|\bar{\chi}_{b,+}^{-1}\right|^2(n_2+1)+\left|\chi_{b,+}^{-1}\right|^2n_2\right) \; ,
\end{align}
%\endgroup
where the response functions are 
\begin{align}
    \chi_{a,\pm}^{-1} &=  -\mathrm{i}(\omega \pm \omega_1)+\frac{\gamma_1}{2} \; , \\
    \chi_{b,\pm}^{-1} &= -\mathrm{i}(\omega \pm \omega_1 - \omega_2)+\frac{\gamma_2}{2} \; , \\
    \bar{\chi}_{b,\pm}^{-1} &= -\mathrm{i}(\omega \pm \omega_1 + \omega_2)+\frac{\gamma_2}{2} \; .
\end{align}
Taking $\displaystyle \int_{-\infty}^{+\infty}\left\langle c[\omega-\omega_1]c[\omega^\prime-\omega_1]\right\rangle\mathrm{d}\omega^\prime $ as an example, Eq.~\ref{eq: example} can be derived as follows:
\begingroup
\allowdisplaybreaks
\begin{align}
    &\int_{-\infty}^{+\infty}\left\langle c[\omega-\omega_1]c[\omega^\prime-\omega_1]\right\rangle\mathrm{d}\omega^\prime \nonumber \\
    =&\int_{-\infty}^{+\infty}\mathrm{d}\omega^\prime\chi_c^2[\omega-\omega_1]\chi_b[\omega-\omega_1]\bar{\chi}_b[\omega-\omega_1]\chi_c^2[\omega^\prime-\omega_1]\chi_b[\omega^\prime-\omega_1]\bar{\chi}_b[\omega^\prime-\omega_1] \nonumber \\
    &\left\{ \gamma_1 \left[ \left( \chi_c^{-1}[\omega-\omega_1]\chi_b^{-1}[\omega-\omega_1] \bar{\chi}_b^{-1}[\omega-\omega_1]+2\mathrm{i}\omega_2 g^2 \right)\left(2\mathrm{i}\omega_2 g^2\right)\left\langle\hat{\xi}_1[\omega-\omega_1]\hat{\xi}_1^\dagger[\omega^\prime-\omega_1]\right\rangle \right.\right.\nonumber \\
    &\left.\left.\,\,\,\,\,\,\,\,\,\,\,\,\,+\left(\chi_c^{-1}[\omega^\prime-\omega_1]\chi_b^{-1}[\omega^\prime-\omega_1] \bar{\chi}_b^{-1}[\omega^\prime-\omega_1]+2\mathrm{i}\omega_2 g^2 \right)\left(2\mathrm{i}\omega_2 g^2\right)\left\langle\hat{\xi}_1^\dagger[\omega-\omega_1]\hat{\xi}_1[\omega^\prime-\omega_1]\right\rangle\right] \right. \nonumber \\
    &\left.\,\,\,\,-g^2\gamma_2\chi_c^{-1}[\omega-\omega_1]\chi_c^{-1}[\omega^\prime-\omega_1] \right.\nonumber \\
    &\left.\,\,\,\,\,\,\,\,\,\left(\bar{\chi}_b^{-1}[\omega-\omega_1]\chi_b^{-1}[\omega^\prime-\omega_1]\left\langle\hat{\xi}_2[\omega-\omega_1]\hat{\xi}_2^\dagger[\omega^\prime-\omega_1]\right\rangle+\chi_b^{-1}[\omega-\omega_1]\bar{\chi}_b^{-1}[\omega^\prime-\omega_1]\left\langle\hat{\xi}_2^\dagger[\omega-\omega_1]\hat{\xi}_2[\omega^\prime-\omega_1]\right\rangle\right)\right\} \nonumber \\
    =&\int_{-\infty}^{+\infty}\mathrm{d}\omega^\prime\chi_c^2[\omega-\omega_1]\chi_b[\omega-\omega_1]\bar{\chi}_b[\omega-\omega_1]\chi_c^2[\omega^\prime-\omega_1]\chi_b[\omega^\prime-\omega_1]\bar{\chi}_b[\omega^\prime-\omega_1] \nonumber \\
    &\Bigg\{ \gamma_1 \Big[ \left( \chi_c^{-1}[\omega-\omega_1]\chi_b^{-1}[\omega-\omega_1] \bar{\chi}_b^{-1}[\omega-\omega_1]+2\mathrm{i}\omega_2 g^2 \right)\left(2\mathrm{i}\omega_2 g^2\right)\left(n_1+1\right)\delta\left[\omega^\prime-\left(2\omega_1-\omega\right)\right]\nonumber \\
    &\,\,\,\,\,\,\,\,\,\,\,\,\,+\left(\chi_c^{-1}[\omega^\prime-\omega_1]\chi_b^{-1}[\omega^\prime-\omega_1] \bar{\chi}_b^{-1}[\omega^\prime-\omega_1]+2\mathrm{i}\omega_2 g^2 \right)\left(2\mathrm{i}\omega_2 g^2\right)n_1\delta\left[\omega^\prime-\left(2\omega_1-\omega\right)\right]\Big]  \nonumber \\
    &\,\,\,\,-g^2\gamma_2\chi_c^{-1}[\omega-\omega_1]\chi_c^{-1}[\omega^\prime-\omega_1] \nonumber \\
    &\,\,\,\,\,\,\,\,\,\left(\bar{\chi}_b^{-1}[\omega-\omega_1]\chi_b^{-1}[\omega^\prime-\omega_1]\left(n_2+1\right)\delta\left[\omega^\prime-\left(2\omega_1-\omega\right)\right]+\chi_b^{-1}[\omega-\omega_1]\bar{\chi}_b^{-1}[\omega^\prime-\omega_1]n_2\delta\left[\omega^\prime-\left(2\omega_1-\omega\right)\right]\right)\Bigg\} \nonumber \\
    =&\chi_c^2[\omega-\omega_1]\chi_b[\omega-\omega_1]\bar{\chi}_b[\omega-\omega_1]\chi_c^2[\omega_1-\omega]\chi_b[\omega_1-\omega]\bar{\chi}_b[\omega_1-\omega] \nonumber \\
    &\Bigg\{ \gamma_1 \Big[ \left( \chi_c^{-1}[\omega-\omega_1]\chi_b^{-1}[\omega-\omega_1] \bar{\chi}_b^{-1}[\omega-\omega_1]+2\mathrm{i}\omega_2 g^2 \right)\left(2\mathrm{i}\omega_2 g^2\right)\left(n_1+1\right)\nonumber \\
    &\,\,\,\,\,\,\,\,\,\,\,\,\,+\left(\chi_c^{-1}[\omega_1-\omega]\chi_b^{-1}[\omega_1-\omega] \bar{\chi}_b^{-1}[\omega_1-\omega]+2\mathrm{i}\omega_2 g^2 \right)\left(2\mathrm{i}\omega_2 g^2\right)n_1\Big]  \nonumber \\
    &\,\,\,\,-g^2\gamma_2\chi_c^{-1}[\omega-\omega_1]\chi_c^{-1}[\omega_1-\omega]\left(\bar{\chi}_b^{-1}[\omega-\omega_1]\chi_b^{-1}[\omega_1-\omega]\left(n_2+1\right)+\chi_b^{-1}[\omega-\omega_1]\bar{\chi}_b^{-1}[\omega_1-\omega]n_2\right)\Bigg\} \nonumber \\
    =&\gamma_1\left|\chi_{a,-}^2\chi_{b,-}\bar{\chi}_{b,-}\right|^2\left(2\mathrm{i}\omega_2g^2\right) \left[\left(\chi_{a,-}^{-1}\chi_{b,-}^{-1}\bar{\chi}_{b,-}^{-1}+2\mathrm{i}\omega_2g^2\right)(n_1+1)+\left(\chi_{a,-}^{*-1}\chi_{b,-}^{*-1}\bar{\chi}_{b,-}^{*-1}+2\mathrm{i}\omega_2g^2\right)n_1\right]\nonumber \\
    &-g^2\gamma_2 \left|\chi_{a,-}\chi_{b,-}\bar{\chi}_{b,-}\right|^2\left(\left|\bar{\chi}_{b,-}^{-1}\right|^2(n_2+1)+\left|\chi_{b,-}^{-1}\right|^2n_2\right) \; ,
\end{align}
\endgroup
Finally, the displacement power spectral density is
\begin{align}
    S_{xx}[\omega]
    &=\frac{\hbar\gamma_1}{2m\omega_1}\left|\chi_{a,-}^2\chi_{b,-}\bar{\chi}_{b,-}\right|^2\left\{ \left|\chi_{a,-}^{-2}\chi_{b,-}^{-1}\bar{\chi}_{b,-}^{-1}+2\mathrm{i}\omega_2g^2\right|^2(n_1+1)+4\omega_2^2g^4n_1 \right.\nonumber \\
    &\quad \quad \left.+2\mathrm{i}\omega_2g^2\left[\left(\chi_{a,-}^{-1}\chi_{b,-}^{-1}\bar{\chi}_{b,-}^{-1}+2\mathrm{i}\omega_2g^2\right)(n_1+1)+\left(\chi_{a,-}^{*-1}\chi_{b,-}^{*-1}\bar{\chi}_{b,-}^{*-1}+2\mathrm{i}\omega_2g^2\right)n_1\right]\right\} \nonumber \\
    &+\frac{\hbar\gamma_1}{2m\omega_1}\left|\chi_{a,+}^2\chi_{b,+}\bar{\chi}_{b,+}\right|^2\left\{ \left|\chi_{a,+}^{-2}\chi_{b,+}^{-1}\bar{\chi}_{b,+}^{-1}-2\mathrm{i}\omega_2g^2\right|^2n_1+4\omega_2^2g^4(n_1+1) \right.\nonumber \\
    &\quad \quad \left.-2\mathrm{i}\omega_2g^2\left[\left(\chi_{a,+}^{*-1}\chi_{b,+}^{*-1}\bar{\chi}_{b,+}^{*-1}-2\mathrm{i}\omega_2g^2\right)(n_1+1)+\left(\chi_{a,+}^{-1}\chi_{b,+}^{-1}\bar{\chi}_{b,+}^{-1}-2\mathrm{i}\omega_2g^2\right)n_1\right]\right\} \; .
\end{align}

For simplicity, we calculate the symmetrized power spectral density 
\begin{align}
    \bar{S}_{xx}[\omega]&=\frac{1}{2}(S_{xx}[\omega]+S_{xx}[-\omega]) \nonumber \\
    &=\frac{\hbar\gamma_1\left(n_1+\frac{1}{2}\right)}{2m\omega_1}\left(\left|\chi_{a,-}+2\mathrm{i}\omega_2g^2\chi_{a,-}^2\chi_{b,-}\bar{\chi}_{b,-}\right|^2+\left|\chi_{a,+}-2\mathrm{i}\omega_2g^2\chi_{a,+}^2\chi_{b,+}\bar{\chi}_{b,+}\right|^2\right) \; .
\end{align}
This $\bar{S}_{xx}[\omega]$ can be separated into two parts $\bar{S}_{xx}[\omega]\triangleq \bar{S}_{xx}^{(+)}[\omega]+\bar{S}_{xx}^{(-)}[\omega]$, where
\begin{align}
    \bar{S}_{xx}^{(+)}[\omega]=\frac{\hbar\gamma_1\left(n_1+\frac{1}{2}\right)}{2m\omega_1}\left|\chi_{a,-}+2\mathrm{i}\omega_2g^2\chi_{a,-}^2\chi_{b,-}\bar{\chi}_{b,-}\right|^2
\end{align}
contains intensity mainly in the positive frequency bands, and
\begin{align}
    \bar{S}_{xx}^{(-)}[\omega]=\frac{\hbar\gamma_1\left(n_1+\frac{1}{2}\right)}{2m\omega_1}\left|\chi_{a,+}-2\mathrm{i}\omega_2g^2\chi_{a,+}^2\chi_{b,+}\bar{\chi}_{b,+}\right|^2
\end{align}
in the negative bands.

\section{Background Intensity}

\begin{figure}
    \centering
    \includegraphics[width=\linewidth]{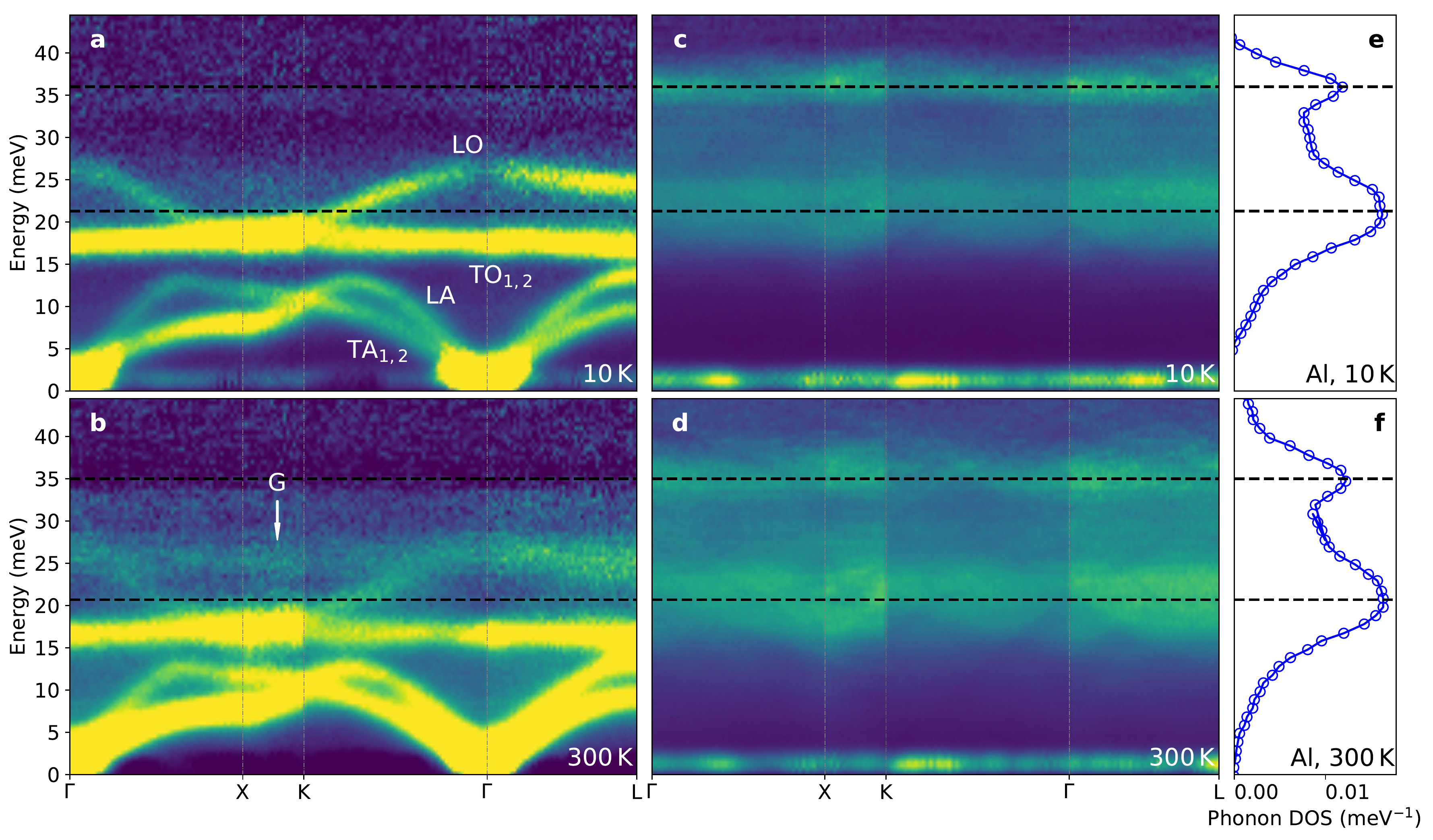}
    \caption{\textbf{Information on background analysis.} \textbf{a-d,} 2D slices through the four-dimensional scattering function  $S(\mathbf{Q}, \varepsilon)$, where $\varepsilon = \hbar \omega$, along  high symmetry lines in the first Brillouin zone, measured at 10\,K (\textbf{a, c}) and 300\,K (\textbf{b, d}) respectively. \textbf{a-b} are the final results of single crystal NaBr and \textbf{c, d} are the background measurements of the empty aluminum can. Corresponding aluminum phonon DOS from previous measurements \cite{Tang2010} are shown in \textbf{e} (10\,K) and \textbf{f} (300\,K). 'G' marks  the intermodulation phonon sideband.} 
    \label{fig: background} 
\end{figure}

When measuring diffuse features in INS,
proper assessment of the background intensity 
from the instrument, and the environment around the sample, is paramount.
Part of the data correction includes 
measurements of the empty sample container under identical 
conditions. 
The empty can background, folded into the first Brillouin zone, is shown in Fig.~\ref{fig: background}c,d, and is compared with Fig.~1 in the manuscript. 
The background has two peaks centered at 20 and 35\,meV (Fig. \ref{fig: background}e,f ), consistent with the 
phonon density of state (DOS) from polycrystalline aluminum 
at 300\,K \cite{Tang2010}. 
The new 
diffuse feature is at 25\,meV, however, so it cannot be the 
residue of the sample container. 
We also confirmed that the diffuse feature 
cannot be formed by the excessive subtraction of the background --
this gives a much wider peak spanning the energy range between 20-35\,meV.
Furthermore, the temperature dependence of the background follows
that of the main dispersions in NaBr, but the diffuse features are
far weaker than the main dispersions at 10\,K, but 
modestly weaker at 300\,K. 

\section{Temperature Dependence of Intermodulation Phonon Sidebands}

\begin{figure}
    \centering
    \includegraphics[width=0.7\linewidth]{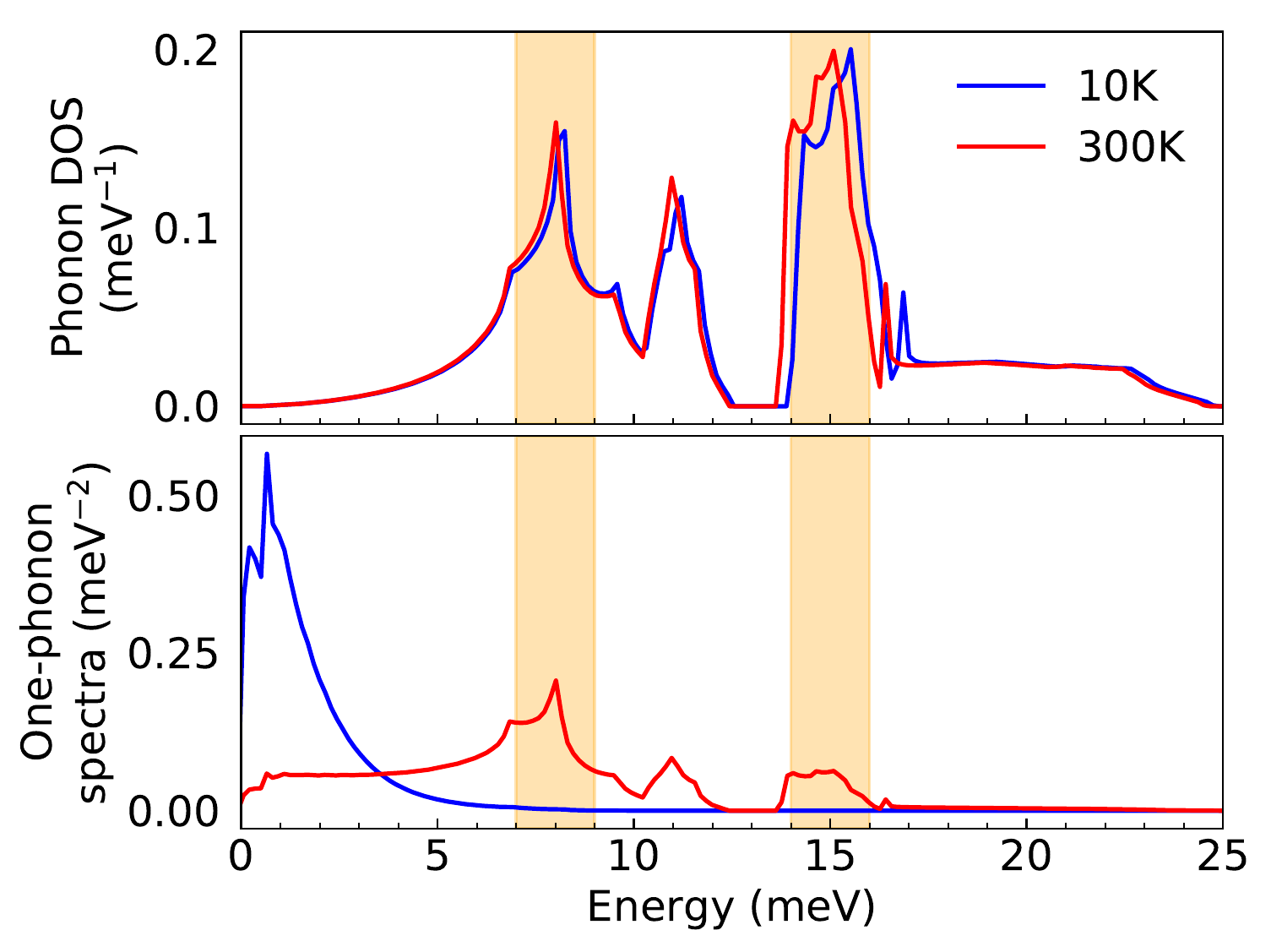}
    \caption{\textbf{Calculated phonon DOS and one-phonon spectra before phonon self-energy corrections.} Phonons involved in the three-phonon processes for IPS feature are indicated by the shaded regions. The one-phonon spectra is given by $A_1(\epsilon)=\frac{g(\varepsilon)}{\varepsilon}\frac{1}{e^{\varepsilon/k_{\rm B}T}-1} $, where $g(\epsilon)$ is the phonon DOS.}
    \label{fig: dosandonephononspectra} 
\end{figure}

The IPS spectra are much 
weaker at 10\,K than at 300\,K,
owing to the temperature dependence of the Planck factors for 
phonon populations. 
%Figure~\ref{fig: dosandonephononspectra} shows how the Planck function 
%makes many more three-phonon processes available at 300\,K
%for generating the ghost phonon branch. 
Figure~\ref{fig: dosandonephononspectra} shows how few one-phonon spectra (before renormalization) are excited at low temperature.
This limits the number of the  TA phonons within 7-9\,meV that 
are available to participate the three-phonon processes to generate the intermodulation sidebands.

\begin{sidewaystable}
\centering
\caption{\footnotesize Fitting parameters.} 
\label{tab: fittingparas}
\begin{tabular}{|c|c||c|c|c|c|c|c|c|c|c|c|}
\hline
\multicolumn{12}{|c|}{\multirow{2}{*}{Peak fitting function: $\displaystyle  y=y_0+\frac{2A}{\pi}\frac{w}{4(x-x_c)^2+w^2}$}}   \\
\multicolumn{12}{|c|}{} \\ \hline
\multicolumn{2}{|c||}{Sub-figure No. in Fig.~\ref{fig: peaks}} & a & b & c & d & e & f & g & h & i & j \\ \hline
\multicolumn{2}{|c||}{q-point}    & [0, 0, 0]  & [0, 0, 0.25]  & [0, 0, 0.5]  & [0, 0, 0.75]  & [0, 0, 1]  & [0, 0.75, 0.75]  & [0, 0.5, 0.5]  &  [0, 0.25, 0.25] & [0.25, 0.25, 0.25]  & [0.5, 0.5, 0.5]  \\ \hline

\multicolumn{2}{|c||}{Offset, $y_0$}     & 0.0013(3)  & -0.00001(6) & -0.0024(2)  & -0.0015(2)  & -0.0019(2)  & -0.0008(2)  & -0.0007(3)  & -0.0037(5)  & -0.0005(3)  & -0.0008(3)  \\ \hline

\multirow{4}{*}{1st. peak}    
& Center, $x_{1}$ & 16.58(7) & 3.86(4)  & 6.05(2)  & 7.36(2)  & 7.73(3)  & 10.33(7)  & 8.48(7)  & 4.48(7)  & 6.09(12)  & 9.01(11)  \\ \cline{2-12}
& Width, $w_1$ & 2.8(3)  & 3.6(2)  & 3.57(11)  & 3.25(11)  & 3.35(11)  & 4.7(3)  & 3.9(3)  & 3.3(3)  & 2.5(4)  & 3.1(4)  \\ \cline{2-12} 
& Area, $A_1$ & 0,.038(4)  & 0.142(9)  & 0.086(3)  & 0.077(3)  & 0.088(3)  & 0.078(5)  & 0.045(5)  & 0.051(10)  & 0.040(9)  & 0.045(6)  \\ \cline{2-12} 
& Height, $H_1$ & 0.010  & 0.025  & 0.013  & 0.014  & 0.015  & 0.010  & 0.007  & 0.006  & 0.010  & 0.009  \\ \hline

\multirow{4}{*}{2nd. peak} 
& Center, $x_{2}$ & 26.3(3)  & 16.74(15)  & 11.45(10)  & 12.37(7)  & 11.92(7)  & 17.90(9)  & 12.18(8)  & 7.4(3)  & 9.1(2)  & 13.52(16)  \\ \cline{2-12} 
& Width, $w_2$ & 3.8(15)  & 3.7(6)  & 4.6(4)  & 2.5(3)  & 2.3(3)  & 4.3(3)  & 2.8(3)  & 6.8(7)  & 4.2(7)  & 2.6(6)  \\ \cline{2-12} 
& Area, $A_2$ & 0.012(5)  & 0.043(8)  & 0.033(3)  & 0.018(2)  & 0.019(2)  & 0.056(4)  & 0.022(3)  & 0.11(2)  & 0.060(12)  & 0.034(10)  \\ \cline{2-12} 
& Height, $H_2$ & 0.002  & 0.007  & 0.005  & 0.005  & 0.005  & 0.008  & 0.005  & 0.010  & 0.009  & 0.008  \\ \hline

\multirow{4}{*}{3rd. peak}    
& Center, $x_{3}$ & -  & 25.9(6)  & 17.07(5)  & 17.40(5)  & 17.27(5)  & 26.3(3)  & 16.90(7)  & 16.74(6)  & 16.68(12)  & 16.40(19)  \\ \cline{2-12} 
& Width, $w_3$ & -  & 7(2)  & 3.6(2)  & 4.0(2)  & 4.8(2)  & 6.6(14)  & 4.2(3)  & 4.4(3)  & 3.2(4)  & 3.0(6)  \\ \cline{2-12} 
& Area, $A_3$  & -  & 0.030(14)  & 0.046(3)  & 0.063(3)  & 0.088(4)  & 0.033(7)  & 0.044(4)  & 0.049(4)  & 0.042(5)  & 0.040(10)  \\ \cline{2-12} 
& Height, $H_3$    & -  & 0.003  & 0.008  & 0.010  & 0.012  & 0.003  & 0.007  & 0.007  & 0.008  & 0.008  \\ \hline

\multirow{4}{*}{4th. peak}
& Center, $x_{4}$ & -  & -  & 24.8(2)  & 26.3(2)  & 26.4(2)  & -  & 21.49(11)  & 25.72(14)  & 25.6(4)  & 25.8(5)  \\ \cline{2-12} 
& Width, $w_4$ & -  & -  & 13.6(16)  & 9.1(11)  & 8.5(12)  & -  & 2.0(5)  & 11.7(10)  & 9.5(19)  & 12(2)  \\ \cline{2-12} 
& Area, $A_4$ & -  & -  & 0.102(14)  & 0.058(7)  & 0.052(7)  & -  & 0.008(2)  & 0.12(2)  & 0.063(13)  & 0.096(19)  \\ \cline{2-12} 
& Height, $H_4$ & -  & -  & 0.005  & 0.004  & 0.004  & -  & 0.003  & 0.007  & 0.004  & 0.005  \\ \hline

\multirow{4}{*}{5th. peak}    
& Center, $x_{5}$ & -  & -  & -  & -  & -  & -  & 25.7(3)  & -  & -  & -  \\ \cline{2-12} 
& Width, $w_5$ & -  & -  & -  & -  & -  & -  & 7.2(12)  & -  & -  & -  \\ \cline{2-12} 
& Area, $A_5$ &  - & -  & -  & -  & -  & -  & 0.038(8)  & -  & -  & -  \\ \cline{2-12} 
& Height, $H_5$ & -  & -  & -  & -  & -  & -  & 0.003  & -  & -  & -  \\ \hline

\end{tabular}
\end{sidewaystable}

%\bibliography{reference_sup.bib}